\documentclass[prb,twocolumn,showpacs,superscriptaddress,floatfix]{revtex4-1}
\usepackage{graphicx,amsfonts,amssymb,amsmath}
\usepackage[colorlinks=true,citecolor=green,linkcolor=red,urlcolor=blue]{hyperref}

\newcommand{\ket}[1]{|#1\rangle}
\newcommand{\bra}[1]{\langle#1|}
\newcommand{\rme}{\mathrm{e}}
\newcommand{\Tr}{\mathrm{Tr}}
\newcommand{\eps}{\varepsilon}

\begin{document}

\title{Finite-temperature mutual information in a simple phase transition}

\author{Johannes Wilms}
\email{johannes.wilms@univie.ac.at}
\address{University of Vienna, Faculty of Physics, Boltzmanngasse 5, 1090 Wien, Austria}

\author{Julien Vidal}
\email{vidal@lptmc.jussieu.fr}
\address{Laboratoire de Physique Th\'eorique de la Mati\`ere Condens\'ee,
CNRS UMR 7600, Universit\'e Pierre et Marie Curie, 4 Place Jussieu, 75252 Paris Cedex 05, France}

\author{Frank Verstraete}
\email{frank.verstraete@univie.ac.at}
\address{University of Vienna, Faculty of Physics, Boltzmanngasse 5, 1090 Wien, Austria}

\author{S\'ebastien Dusuel}
\email{sdusuel@gmail.com}
\address{Lyc\'ee Saint-Louis, 44 Boulevard Saint-Michel, 75006 Paris, France}


\begin{abstract}

We study the finite-temperature behavior of the Lipkin-Meshkov-Glick model
with a focus on correlation properties as measured by the mutual information.
The latter, which quantifies the amount of both classical and quantum
correlations, is computed exactly in the two limiting cases of vanishing
magnetic field and vanishing temperature.
For all other situations , numerical results provide evidence of a finite mutual information at all temperatures except at criticality. There, it
diverges as the logarithm of the system size, with a prefactor that can take
only two values, depending on whether the critical temperature vanishes or not.
Our work provides a simple example in which the mutual information appears as a
powerful tool to detect finite-temperature phase transitions, contrary to
entanglement measures such as the concurrence.

\end{abstract}

\pacs{03.65.Ud, 03.67.-a,89.70.Cf,05.70.Fh}

\maketitle

\section{Introduction}

The theory of entanglement, originally developed in the field of
quantum information theory, has recently become a very valuable 
tool to describe the phase diagram and the properties of strongly-correlated quantum many-body systems \cite{Eisert10}. For example, the
study of the scaling of the entanglement entropy as a function of the
system size has become the standard tool to detect the central charge
in critical quantum spin chains \cite{Holzhey94,GVidal03}, and the entanglement spectrum in
topological insulators is providing a very nice characterization of
the spectrum of the corresponding edge modes \cite{Fidkowski10,Turner10}.
Also, second-order quantum phase transitions are characterized by the emergence of a
large amount of entanglement.

Almost all of those results were obtained in the zero-temperature
regime, {\it i.e.}, for ground states. In that regime, all correlations are
quantum correlations, and the subtleties of mixed-state quantum
entanglement are avoided. From the point of view of quantum systems at
finite temperature, interesting questions about separability, entanglement
cost and entanglement of distillation could in principle be posed, but
those concepts do not seem to have any clear operational meaning in
the context of a quantum many-body system as there is no clear
separation of the different degrees of freedom in that case. Indeed, at finite temperature, it  seems to make much more sense to
look at the total correlations in the system, without trying to
distinguish classical from quantum degrees of freedom: it is precisely
the fact that (quasi-) long range correlations emerge in the many-body
system that makes them so interesting, and whether those correlations
are quantum or classical is a question that would anyway depend on the
way the system is partitioned and on the definition of
entanglement.

The most appealing quantity that measures the total amount of
correlations in a quantum system is the mutual information. For pure
(zero-temperature) states it reduces to twice the entanglement entropy, and effectively quantifies the amount of (Shannon) information we acquire about the configuration of part of
the system by measuring another part. For instance, in the case of an Ising
ferromagnet at zero temperature, the mutual information
between two parts of the system is one bit, as the measurement of the
spins in one part reveals one bit of information (up or down) about
the configuration of the spins in the other part. In such purely classical cases, the behavior of the mutual information as a function of the temperature and the system size is very intriguing and has
been discussed in Refs.~\onlinecite{Arnold96,Sole96,Matsuda96,Wilms11,Nussinov12}. 
The really nice features of the mutual information are that:  ($i$) it provides a natural way of extending
the study of entanglement entropy to the case of finite temperature;
($ii$) it has a nice operational interpretation; ($iii$) it provides
a measure of all correlations in a quantum many-body system, not
just the two- or three-body correlations. This last point is of special
importance when studying collective models such as the
Lipkin-Meshkov-Glick (LMG) model \cite{Lipkin65,Meshkov65,Glick65}, where there is no clear notion of
locality.

The goal of this paper is to initiate the study of the behavior of
the mutual information in quantum many-body systems at finite
temperature following recent studies \cite{Hastings10,Melko10,Singh11,Kallin11}. 
Questions that will be addressed include, {\it e.g.}, the study of
how the scaling of the mutual information depends on the temperature,
and how precisely the mutual information allows one to identify  phase
transitions. We will focus our attention on the LMG
model, because it is a quantum many-body system that exhibits a
nontrivial phase transition with true many-body correlations, while still
being simple enough such that the calculation of the mutual
information is tractable.

We will show that the mutual information yields precise information about the location of the phase transition. An intriguing difference in its finite-size scaling is obtained between the zero versus non-zero temperature case.

\section{Model}
\label{sec:model}

\subsection{Hamiltonian and symmetries}
\label{subsec:symm}

Let us consider a system of $N$ fully-connected spins 1/2 whose Hamiltonian
reads
%
%
\begin{equation}
\label{eq:ham}
H=-\frac{S_x^2}{N}-h\,S_z, 
\end{equation}
%
%
where $S_\alpha=\tfrac{1}{2}\sum_{j=1}^N\sigma_j^\alpha$ is the total-spin
operator along the $\alpha=x$, $y$, $z$ direction, $\sigma_j^\alpha$ being the
usual Pauli matrix at site $j$. Without loss of generality, we suppose that
$h \geqslant 0$. 
The Hamiltonian (\ref{eq:ham}) can be seen as a ferromagnetic transverse-field
Ising model with infinite-range interactions (see Ref.~\onlinecite{Dutta11} for
a recent review). It is also a special case of the LMG model
\cite{Lipkin65,Meshkov65,Glick65} whose ground-state entanglement properties
have triggered much attention during the last years
\cite{Vidal04_1,Dusuel04_3,Dusuel05_2,Latorre05_2,Barthel06_2,Vidal07,
Orus08_2,Wichterich10}.

As can be seen easily, this Hamiltonian preserves the total spin, {\it i.e.}, $[H,\mathbf{S}^2]=0$. This conservation rule, together
with the so-called spin-flip symmetry $[H,\prod_j\sigma_j^z]=0$, allows one to
investigate this model both analytically and numerically.
Recently, the exact spectrum has been obtained in the thermodynamical limit, in
the maximum spin sector \mbox{$s=N/2$} where the ground state lies
\cite{Ribeiro07,Ribeiro08}.
Unfortunately, a complete derivation of the full spectrum (in arbitrary $s$ sectors) that is crucial for the understanding of the finite-temperature properties is still missing.

%
%
\subsection{Phase diagram}
\label{subsec:diagram}
%
%

In order to get a first feeling about the model, let us start with a discussion
of the finite-temperature phase diagram. It can be obtained using a standard
mean-field approach (see for instance Refs.~\onlinecite{Dutta11,Quan09}) that we briefly describe here. 
We introduce the order parameter (proportional to the average magnetization along the
$x$-direction) $m_x = \langle \sigma_j^x \rangle = \tfrac{2}{N}\langle S_x\rangle$,
and write that $\sigma_j^x = m_x + (\sigma_j^x-m_x)$.
Then one can plug this expression into the Hamiltonian, neglect fluctuations
involving $(\sigma_j^x-m_x)^2$ and keep only dominant terms in a 
$1/N$-expansion. This yields an effective Hamiltonian of a large spin
$\boldsymbol{S} = (S_x,S_y,S_z)$ in an effective magnetic field
$\boldsymbol{h}_\mathrm{eff} = (m_x, 0, h)$, that reads
%
%
\begin{equation}
\label{eq:ham_eff}
H_\mathrm{eff} = -\boldsymbol{h}_\mathrm{eff}\cdot\boldsymbol{S},
\end{equation}
%
%
up to an unimportant constant shift of $N \frac{m_x^2}{4}$.
One still has to ensure that $m_x$ coincides with the thermodynamical
average $\tfrac{2}{N}\langle S_x\rangle$. Therefore, one needs the partition
function
%
%
\begin{equation}
\label{eq:Zeff}
Z_\mathrm{eff} = \mathrm{Tr}\left(\mathrm{e}^{-\beta H_\mathrm{eff}}\right)
= \left[2\cosh\left(\frac{\beta\sqrt{m_x^2+h^2}}{2}\right)\right]^N,
\end{equation}
%
%
where we have used the inverse temperature $\beta=1/T$. Noticing that
%
%
\begin{equation}
m_x = \frac{1}{Z_\mathrm{eff}}
\mathrm{Tr}\left(\frac{2S_x}{N}\mathrm{e}^{-\beta H_\mathrm{eff}}\right)
= \frac{2}{N\beta}\frac{\partial \ln Z_\mathrm{eff}}{\partial m_x},
\end{equation}
%
%
one obtains that  either $m_x=0$ or $m_x$ satisfies
%
%
\begin{equation}
\label{eq:m}
\sqrt{m_x^2+h^2}=\tanh \left(\frac{\sqrt{m_x^2+h^2}}{ 2 \, T}\right). 
\end{equation}
%
%
The above equation has a solution only for $T<T_\mathrm{c}(h)$, where the
critical temperature reads
%
%
\begin{equation}
\label{eq:Tc}
T_\mathrm{c}(h)=\frac{h}{2\tanh^{-1}(h)}.
\end{equation}
%
%
%
%
%
\begin{figure}[t]
\includegraphics[width=0.8\columnwidth]{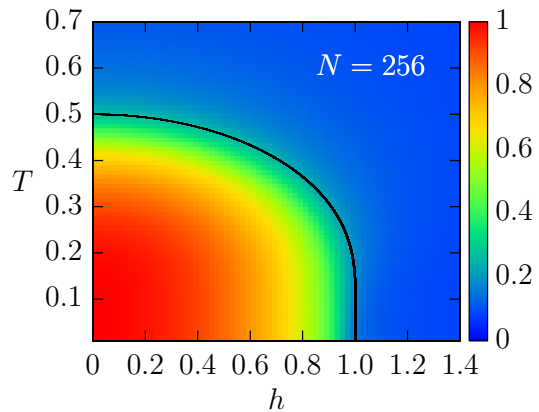}
\caption{Density plot of the order parameter $m$ in the $(h,T)$ plane for a
finite-size system ($N=256$).
The black line indicates the transition line $T_{\rm c}(h)$ whose expression is
given in Eq.~(\ref{eq:Tc}).}
\label{fig:density_op}
\end{figure}
%
%
%
%
%
\begin{figure}[b]
\includegraphics[width=0.8\columnwidth]{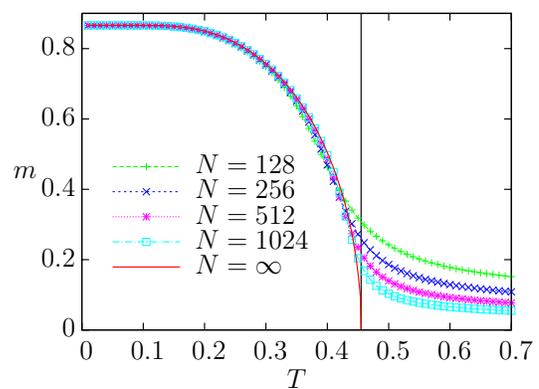}
\caption{Order parameter $m$ for $h=1/2$ as a function of temperature for
different values of $N$. The critical exponent is $\beta_\mathrm{MF}=1/2$.
The black line indicates the transition temperature
$T_{\rm c}(1/2)\simeq 0.46$.}
\label{fig:op_T}
\end{figure}
%
%
%
This critical temperature, which is always smaller than $1/2$, is represented
as a black line in Fig.~\ref{fig:density_op}.
We thus find, as expected, a disordered (paramagnetic) phase at high
temperature $T>T_{\rm c}(h)$ with a vanishing order parameter $m_x$ and
an ordered (ferromagnetic) phase for $T<T_{\rm c}(h)$ where $m_x$ acquires a
finite value. One can also easily show that, in the vicinity of $T_{\rm c}(h)$, the order parameter vanishes as
$m_x \sim (T_\mathrm{c}-T)^{\beta_\mathrm{MF}}$ with the standard mean-field
exponent $\beta_\mathrm{MF}=1/2$ (see for instance Ref.~\onlinecite{Botet83}).

The validity of the mean-field approach can be checked numerically, provided one
considers the alternative quantity $m=\tfrac{2}{N}\sqrt{\langle S_x^2 \rangle}$.
Indeed, there is no symmetry breaking for any finite value of $N$, so that $m_x$
always vanishes in numerical simulations. As the system is fully connected, one
furthermore expects that $m^2$ and $m_x^2$ become equal in the thermodynamical
limit. Exact diagonalization results are compared  with the mean-field
prediction in Figs.~\ref{fig:density_op} and \ref{fig:op_T} where an excellent agreement can be observed.\\\\\\\\

%
%
\begin{figure}[t]
\includegraphics[width=0.8\columnwidth]{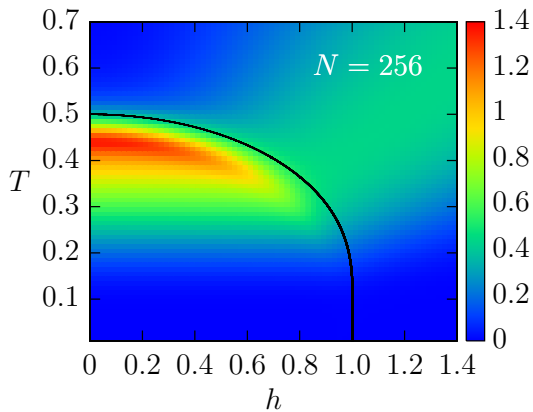}
\caption{Density plot of the heat capacity per spin $c$ in the $(h,T)$ plane
for a finite-size system ($N=256$).
The black line indicates the transition line $T_{\rm c}(h)$.}
\label{fig:density_hc}
\end{figure}
%
%
%
%
%
\begin{figure}[b]
\includegraphics[width=0.8\columnwidth]{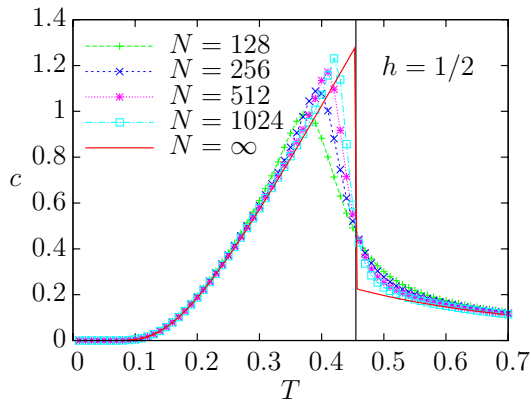}
\caption{Heat capacity per spin $c$ for $h=1/2$ as a function of temperature
for different values of $N$. The discontinuity at $T_{\rm c}(1/2)\simeq 0.46$
(black vertical line) is associated with the critical mean-field exponent
$\alpha_\mathrm{MF}=0$.}
\label{fig:hc_T}
\end{figure}
%
%

Along the same line, one can  compute the heat capa\-city defined as
$C=\frac{\partial U}{\partial T}$, where $U$ is the average energy
%
%
\begin{equation}
U = \frac{1}{Z_\mathrm{eff}}
\mathrm{Tr}\left(H_\mathrm{eff}\mathrm{e}^{-\beta H_\mathrm{eff}}\right)
= -\frac{\partial \ln Z_\mathrm{eff}}{\partial \beta}.
\end{equation}
%
%
%
The heat capacity per spin thus reads
%
%
\begin{equation}
c = \frac{C}{N}
= \frac{\beta^2}{N}\frac{\partial^2 \ln Z_\mathrm{eff}}{\partial \beta^2},
\end{equation}
%
%
which can be straightforwardly evaluated using Eq.~(\ref{eq:Zeff}).
As can be seen in Figs.~\ref{fig:density_hc} and \ref{fig:hc_T}, analytical results stemming from the mean-field approach are still perfectly consistent with the numerical data. Note that one recovers the
(well-known) fact that the heat capacity displays a jump at the transition, so
that $C \sim |T-T_{\rm c}|^{- \alpha_\mathrm{MF}}$ with $\alpha_\mathrm{MF}=0$.

%
%
\subsection{Finite-size corrections}
\label{subsec:finitesize}
%
%

Mean-field results obtained in Sec.~\ref{subsec:diagram} are valid in the
thermodynamical limit \mbox{($N\to \infty$)}.
However, it is interesting to pay attention to finite-size corrections that
contain valuable informations. These can be studied quantitatively since,
contrary to usual quantum spin systems, the symmetries of the LMG model
discussed in Sec.~\ref{subsec:symm} allow one to perform numerical computations for relatively
large system sizes (at least $N\sim 10^3$ for all quantities considered in
this work).

\begin{figure}[htbp]
\includegraphics[width=\columnwidth]{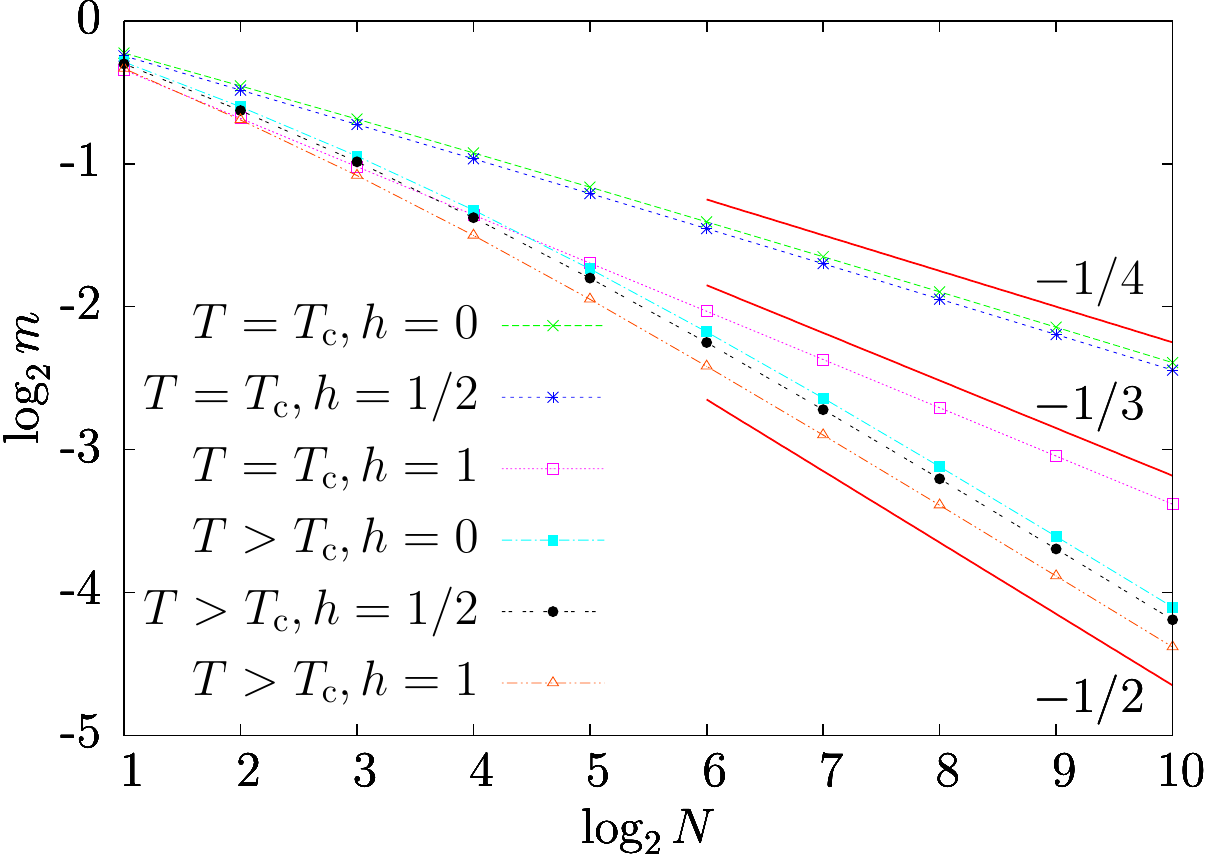}
\caption{${\rm Log}_2$-$\log_2$ plot of $m$ {\it vs} $N$  for $h=0$, $1/2$, and $1$
computed at the critical temperature $T_{\rm c}(h)$ and at high temperature
$T=0.7$ where $m_x=0$. Red lines are straight lines with slopes $-1/2$, $-1/3$, and $-1/4$
(see the text). }
\label{fig:op_N}
\end{figure}

At zero temperature, these finite-size corrections have been shown to be very
sensitive to the transition \cite{Botet83,Dusuel04_3,Dusuel05_2,Liberti10}. For
instance, corrections to the thermodynamical value of the order parameter $m$
behave as $N^{-1/3}$ at the critical point and as $N^{-1/2}$ otherwise. 
At finite temperature, we found numerical evidence for a similar behavior for
$m$ away from criticality [more precisely, $m^2=m_x^2+a/N+O(1/N^2)$] but, on the critical line $T_{\rm c}(h)>0$, $m$ rather
seems to vanish as $N^{-1/4}$ as can be seen in Fig.~\ref{fig:op_N}. 
Unfortunately,  methods developed to analyze the zero-temperature problem
\cite{Dusuel04_3,Dusuel05_2} cannot be used at finite temperature. This is
mainly due to the fact that, for $T\neq 0$,  one needs to consider all spin $s$
sectors and not only the maximum $s$ sector where the ground state lies.
However, for $h=0$ (no quantum fluctuations),  one can compute the leading finite-size corrections 
of all quantities at and away from criticality ($T_{\rm c}(0)=1/2$). In this special case, as shown in Sec.~\ref{subsec:classical}, 
$m$ is found to vanish as $N^{-1/4}$ giving a first indication that the nontrivial finite-size behavior
at $T_{\rm c}(h)>0$ might be inferred from the zero-field problem.
Although one can argue that quantum fluctuations are irrelevant for the study of critical behavior at  finite-temperature, it is striking to recover these behaviors from the trivial $h=0$ case.
%
%
%
%

%
%
\section{Mutual information}
\label{sec:entanglement} 
%
%
\subsection{Generalities}
\label{subsec:entanglement_generalities}

We now turn to the main focus of the paper, namely the analysis of the
finite-temperature correlations. As already mentioned, ground-state
entanglement properties relevant for the zero-temperature problem have been
studied extensively \cite{Vidal04_1,Dusuel04_3,Latorre05_2,Dusuel05_2,
Barthel06_2,Vidal07,Orus08_2,Wichterich10}. Note also that several quantities based on fidelity have also been proposed to capture the zero-temperature transition in the LMG model \cite{Kwok08,Ma08} but they do not measure entanglement properties.
By contrast, there have only been a few investigations concerning entanglement in the LMG model at finite temperature \cite{Canosa07,Matera08} mainly because entanglement measures for mixed states are rare and hardly computable.

\begin{figure}[h]
\includegraphics[width=0.8\columnwidth]{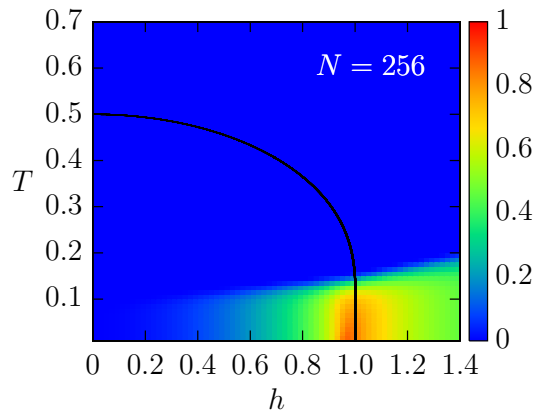}
\caption{Density plot of the concurrence in the $(h,T)$ plane for a finite-size
system ($N=256$).
The black line indicates the transition line $T_{\rm c}(h)$.}
\label{fig:density_concurrence}
\end{figure}

This last decade, many studies have shown that entanglement measures could be
used to detect quantum (zero-temperature) phase transitions (see for instance
Ref.~\onlinecite{Amico08} for a review). Here, following recent works in
two-dimensional quantum spin systems \cite{Hastings10,Melko10,Singh11}, we wish
to investigate the same problem at finite temperature. Concerning the LMG model,
the concurrence \cite{Wootters98} (which characterizes the entanglement between
two spins in an arbitrary state) is up to now the only entanglement measure that
has been computed both at zero \cite{Vidal04_1,Dusuel04_3, Dusuel05_2} and at
finite temperature \cite{Matera08}. 
However, as can be seen in Fig.~\ref{fig:density_concurrence}, this quantity is
quite surprisingly insensitive to the transition line at finite temperature.
%
%
%
%

Alternatively, one may consider the negativity \cite{Vidal02} that is also a
good measure for mixed states and that is not restricted to two-spin
correlations. This measure has been successfully used to study the quantum phase
transition in the LMG model \cite{Wichterich10} and other fully-connected models \cite{Filippone11}. It is however already difficult to compute at zero
temperature, so one can expect troubles to obtain it at finite temperature for
systems of decent sizes (analytical expressions in the thermodynamical limit
being out of reach).
To our knowledge, apart from these two measures (concurrence and negativity),
there exists only one reliable quantity susceptible to be used for mixed
states, namely the mutual information, whose definition is based on
entanglement entropy.

To obtain informations about many-body correlations of a system, the
entanglement entropy is usually the measure of choice for pure states. It is
defined as the entropy of the reduced density matrix $\rho_A$ obtained, after partitioning the system into two parts $A$ and $B$, by tracing out the density matrix $\rho$ over subsystem $B$~: $\rho_A=\mathrm{Tr}_B\rho$. For pure states, the same entropy is obtained if one exchanges $A$ and $B$, although the
result obviously depends on the bipartition.
For mixed states, however, the entanglement entropy is dominated by a
contribution related to the total entropy of the system under consideration
(which vanishes for pure states).
Therefore, the natural thing to do is to subtract this total entropy in a
suitable way. This leads to the following definition of the mutual information
%
%
\begin{equation}
\mathcal{I}(A,B)=\mathcal{E}_A+\mathcal{E}_B-\mathcal{E}_{A B},
\label{eq:mutinf_def}
\end{equation}
%
%
where the von Neumann entropy $\mathcal{E}_A=-\mathrm{Tr}(\rho_A\log_2\rho_A)$
is the quantum generalization of the Shannon entropy. The Shannon entropy
provides an asymptotic description of the properties of a system (a probability
distribution of states), in the sense that it describes the average number of
bits needed to encode a state of the system. This average is taken over all
the states of the system, using their respective probabilities of occurrence.
In other words, Shannon entropy corresponds to the average amount of randomness that can be extracted from the system.  Note that it is also possible to work with R\'enyi entropies defined as 
\mbox{$\mathcal{E}_A(\kappa)=\frac{1}{1-\kappa}\log_2\mathrm{Tr}\rho_A^\kappa$}\cite{Koenig09}, 
but let us simply underline that when $\kappa \to 1$, they become identical to the aforementioned Shannon/von Neumann version that is considered in the following.

Before discussing the results, it is important to understand that mutual
information is \emph{not} a measure for the ``quantumness'' of a system: it
captures both classical and quantum correlations alike.
To make the distinction between both, one
could rather study the so-called quantum discord \cite{Ollivier01,Sarandy09}. However, such
a measure is clearly artificial as the quantum discord of two states
arbitrarily close to each other can vanish for one of them and be
positive for the other one \cite{Ferraro09}. Such mixed-state entanglement  measures were
constructed from the point of view of delocalized systems, and do not
necessarily make sense in the context of condensed-matter systems.

\begin{figure}[t]
\includegraphics[width=0.8\columnwidth]{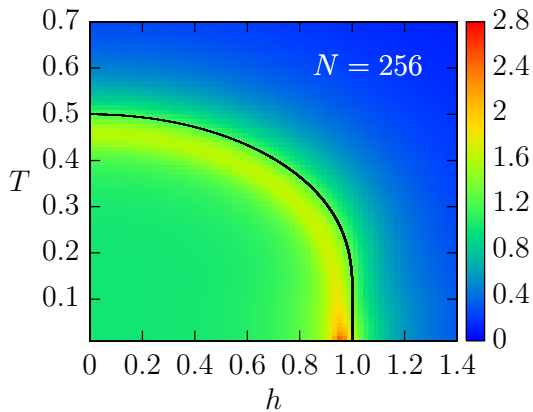}
\caption{Density plot of the mutual information $\mathcal{I}$ in the $(h,T)$
plane for a finite-size system ($N=256$).
The black line indicates the transition line $T_{\rm c}(h)$.}
\label{fig:density_mi}
\end{figure}

In the present work, we shall not try to distinguish between quantum and classical
correlations and will stick to computing the mutual information. Indeed, such a quantity should be sensitive to strong fluctuations of the many-body correlations arising at a phase transition.
Numerical results are displayed in Fig.~\ref{fig:density_mi}. Following
a ray in the $(h,T)$ plane of this figure, one gets that
$\mathcal{I}(A,B)$ is almost equal to $1$ near the origin, grows as one goes
away from the origin, reaches a maximum before transition line, and then
decreases to zero. Assuming that the maximum is located at the transition line
in the thermodynamical limit (see Sec.~\ref{subsec:results} for a confirmation
of this), one might thus argue that $\mathcal{I}(A,B)$ detects the phase
transition at least as well as usual thermodynamic quantities such as the
order parameter or the heat capacity discussed in Sec.~\ref{subsec:diagram}.
However, one should keep in mind that computing the mutual information is not as easy as computing an order parameter, even in the simplest case $(h=0)$ that we shall now discuss.

%
%
%
%

%
%
%
\subsection{Exact results for $h=0$}
\label{subsec:classical} 
%
%

Let us consider the classical (quantum fluctuation free) case $h=0$ for which
the Hamiltonian (\ref{eq:ham}) simplifies to
%
%
\begin{equation}
    H = -\frac{S_x^2}{N}.
\end{equation}
%
%
Its eigenstates are those of $S_x$ and can be chosen as separable states
$|p,i\rangle$, with $p$ spins pointing in the $-x$ direction and
$(N-p)$ spins pointing in the $+x$ direction. The variable $i$ allows to
distinguish between the $\left(\begin{array}{c}N\\p\end{array}\right)$ such states
which are degenerate and have eigenenergy
$E_p=-N\left(\tfrac{p}{N}-\tfrac{1}{2}\right)^2$.

To obtain the mutual information, one first needs to compute the total entropy
of the system at finite temperature and, hence, the partition function 
%
%
\begin{equation}
\label{eq:Z_zero_field}
Z(\beta)=\Tr \: {\rm e}^{-\beta H}=\sum_{p=0}^{N}
 \left(
\begin{array}{c}
N
\\
p
\end{array}
\right){\rm e}^{-\beta E_p},
\end{equation}
%
%
where as before, $\beta=1/T$ denotes the inverse temperature.
Despite the apparent simplicity of $H$, it is difficult to compute $Z$
analytically for arbitrary finite $N$.
Thus, in the following, we shall focus on the physically relevant
large-$N$ limit to analyze the thermodynamical limit and its neighborhood.
Computations get easier in this limit because the discrete sum in (\ref{eq:Z_zero_field}) can be
approximated by an integral with, as is well-known, an error of the order
$1/N^2$. This is fortunate since, in order to compute the mutual
entropy in the present problem, we will need to compute
the first correction to the infinite $N$ limit.

The very first step in the calculation is  to perform a large-$N$ expansion
of the binomials
%
%
\begin{equation}
\label{eq:bin}
\left(
\begin{array}{c}
N
\\
p
\end{array}
\right)
= \frac{N!}{p!\:(N-p)!}=\frac{\Gamma(N+1)}{\Gamma(p+1) \Gamma(N-p+1)}.
\end{equation}
%
%
Using the series expansion of the Euler-$\Gamma$ function at large $N$ (or the
Stirling formula) and skipping terms of relative order $1/N^2$, one gets 

%
%
\begin{eqnarray}
\label{eq:bin_approx}
\left(
\begin{array}{c}
N
\\
p
\end{array}
\right)
&=& \sqrt{\frac{2}{\pi N}} \frac{1}{\sqrt{1-4 \eps^2}} \times \nonumber\\
&& {\rm e}^{-N
\left[\left(\frac{1}{2}-\eps\right)\log\left((\frac{1}{2}-\eps\right)
+\left(\frac{1}{2}+\eps\right)\log\left(\frac{1}{2}+\eps\right)\right]}
\times \nonumber\\
&&\left[1-\frac{1}{N} \frac{3+4\eps^2}{12(1-4\eps^2)}+O\left(1/N^2\right)\right],
\end{eqnarray}
%
%
where we set $\eps=\frac{p}{N}-\frac{1}{2}$. Then, replacing the sum by an integral in
Eq.~(\ref{eq:Z_zero_field}) gives

%
%
\begin{eqnarray}
Z(\beta)&=& 2^N \sqrt{\frac{2N }{\pi}} \int_{-\frac{1}{2}}^{+\frac{1}{2}}
\frac{{\rm d} \eps}{\sqrt{1-4 \eps^2}} \times \nonumber\\
&&
\left[1-\frac{1}{N} \frac{3+4\eps^2}{12(1-4\eps^2)}+O\left(1/N^2\right)\right]
\mathrm{e}^{-N \varphi(\eps)},\quad 
\end{eqnarray}
with
\begin{eqnarray}\varphi(\eps) &=& \left(\frac{1}{2}-\eps\right)\log\left(\frac{1}{2}-\eps\right)
+\left(\frac{1}{2}+\eps\right)\log\left(\frac{1}{2}+\eps\right)\nonumber\\
&&\qquad\qquad+\log 2 -\beta \eps^2.
\end{eqnarray}
%
%
Furthermore, the exponential term $\mathrm{e}^{-N \varphi(\eps)}$ allows us to
extend the integration range to $\mathbb{R}$. Indeed, the effect of this extension consists in  exponentially small terms $(\propto \mathrm{e}^{-\alpha N}$) to which we anyway do not have access.
The resulting integrals can be evaluated using the standard Laplace's method
(also known as saddle-point or stationary-phase approximation) though one has
to take care of computing the subleading corrections \cite{Bender99}.

At and above the critical temperature
($\beta \leqslant \beta_\mathrm{c}(0)=2$), the minimum of $\varphi(\eps)$
is found for $\eps=0$ and the result of the saddle-point approximation reads
%
%
\begin{equation}
\label{eq:Z_highT}
Z(\beta<2) = 2^N \sqrt{\frac{2 }{2-\beta}} \left[1-\frac{1}{N}\frac{\beta^2}{4(2-\beta)^2}+O(1/N^2)\right],
\end{equation}
%
%
and 
%
%
\begin{eqnarray}
Z(\beta=2)&=& 2^{N-1}  \frac{3^{1/4} N^{1/4}\Gamma(1/4)}{\sqrt{\pi}}
\bigg[1+\frac{2 \sqrt{3} \:\Gamma(3/4) }{5\sqrt{N}\:\Gamma(1/4)} \nonumber \\
&&  -\frac{1}{280 N} -\frac{\Gamma(3/4) }{20\sqrt{3} N^{3/2}\:\Gamma(1/4)} +O(1/N^2)\bigg]. \nonumber \\
&&
\end{eqnarray}
%
%
Note that, in this approach, $Z(\beta=2)$ does not coincide with $\lim_{\beta
\to 2} Z(\beta<2)$ (which is divergent) since the integrals stemming from the
stationary-phase approximation is not gaussian anymore at criticality.
Furthermore, we see that $Z(\beta=2)/2^N$ diverges with a non-trivial
finite-size scaling exponent, as $N^{1/4}$. This result, which is obtained by
directly computing $Z(\beta=2)$, is consistent with Eq.~(\ref{eq:Z_highT}).
Indeed, adapting the finite-size scaling argument developed in
Refs.~\onlinecite{Dusuel04_3,Dusuel05_2} for the zero-temperature problem,
one can write (for $\beta$ smaller than $2$ but close to 2)
$Z(\beta\lesssim 2)/2^N=z_\infty f[N(2-\beta)^2]$ where
$z_\infty=\sqrt{2/(\beta-2)}$ is
the thermodynamical limit value of $Z(\beta<2)/2^N$, and $f$ is a scaling
function. As this quantity cannot be singular for  finite $N$, one must have
$f(x)\sim x^{1/4}$ so that the singularity at $\beta=2$ disappears~:
$Z(\beta\lesssim 2)/2^N\sim (\beta-2)^{-1/2} [N(2-\beta)^2]^{1/4}\sim N^{1/4}$.

In the low-temperature phase ($\beta >2$), $\varphi(\eps)$ has two symmetric
minima the position of which can only be computed numerically. Once determined, one can still perform the
Gaussian integral resulting from the second-order Taylor expansion around these
minima, which yields a ``numerically exact" result in the infinite-$N$ limit.
As a consequence, we only give analytical expressions for $\beta \leqslant 2$  but we also show the exact numerical results for $\beta >2$ in the figures.

To compute the entropy in the thermodynamical limit, one needs to compute the internal energy of the system using the same approximation (\ref{eq:bin_approx}) of the binomials. This quantity is defined as
%
%
\begin{equation}
U(\beta)= \Tr(\rho H)=-\frac{\partial \ln Z(\beta)}{\partial \beta},
\end{equation}
%
%
where $\rho=\tfrac{1}{Z}\mathrm{e}^{-\beta H}$ is the thermal density matrix.
For $\beta < 2$, it can be obtained from Eq.~(\ref{eq:Z_highT}) but, as
previously, special care must be taken to deal with the critical case for
which one can show that
%
%
\begin{widetext}
\begin{eqnarray}
U(\beta=2) &=&-\frac{\sqrt{3 N}\: \Gamma(3/4)}{2 \Gamma(1/4)}\bigg[1-\frac{2 \sqrt{3} \:\Gamma(3/4) }{5\sqrt{N}\:\Gamma(1/4)} 
+1 2\frac{\Gamma(1/4)^2+7\:\Gamma(3/4)^2}{175 N \: \Gamma(1/4)^2} \nonumber \\
&& -\frac{10 \: \Gamma(1/4)^4 +32 \: \Gamma(1/4)^2\: \Gamma(3/4)^2+504\:
\Gamma(3/4)^4}{875\sqrt{3} N^{3/2}\:\Gamma(1/4)^3 \Gamma(3/4)}+O(1/N^2)\bigg],
\end{eqnarray}
\end{widetext}
%
%
so that the internal energy diverges as $N^{1/2}$.

Given that,  $U(\beta)=\langle H \rangle=-\frac{2}{N} \langle S_x^2 \rangle$, this latter result directly implies that the order parameter
\mbox{$m=\frac{2}{N}\sqrt{\langle S_x^2 \rangle}$} vanishes at criticality as
$N^{-1/4}$, as was already noticed in Fig.~\ref{fig:op_N}.

The entropy can finally be computed (analytically for $\beta \leqslant 2$ and
numerically for $\beta > 2$) since
%
%
\begin{equation}
\mathcal{E}_{A B}(\beta)=-\Tr( \rho \log_2 \rho)=\log_2 Z(\beta)
+\frac{\beta U(\beta)}{\log 2}. 
\label{eq:entrop}
\end{equation}
%
%

Computing the mutual information is another game, the most difficult part of
which lies in the derivation of the large-$N$ behavior of the partial entropy
%
%
\begin{equation}
\mathcal{E}_{A}(\beta)=- \Tr( \rho_A \log_2 \rho_A),
\end{equation}
%
%
where $ \rho_A =\Tr_B \rho$ is the reduced density matrix. It is thus mandatory
to find an expression for $\rho_A$. With this aim in mind, let us split the
system into two parts $A$ and $B$ containing $N_A$ and $N_B=N-N_A$ spins
respectively. Next, let us decompose the eigenstates of $H$ as
\mbox{$|p,i\rangle= |p_A,i_A\rangle_A \otimes |p_B,i_B\rangle_B$} with
$p_A+p_B=p$ and where $|p_j,i_j\rangle_j$ denotes a state of subsystem $j=A,B$
that has $p_j$ spins pointing in the $-x$ direction. Variable
$i_j$ can take $\left(\begin{array}{c}N_j\\p_j\end{array}\right)$  values and
depend on which spins point in the $-x$ direction.
This decomposition allows one to write the reduced density
matrix as

%
%
\begin{eqnarray}
\rho_A &=&\frac{1}{Z}\Tr_B \sum_{p,i}
\mathrm{e}^{N\beta\left(\frac{p}{N}-\frac{1}{2}\right)^2}
|p,i\rangle \langle p,i| ,\\
&=&\frac{1}{Z}\Tr_B \sum_{p_A,i_A,p_B,i_B}
\mathrm{e}^{N\beta\left(\frac{p_A}{N}+\frac{p_B}{N}-\frac{1}{2}\right)^2}
\times\\
&&
\quad|p_A,i_A\rangle_A \otimes |p_B,i_B\rangle_B\,
\mbox{}_A\langle p_A,i_A | \otimes \mbox{}_B\langle p_B,i_B|, \nonumber \\
&=&\sum_{p_A,i_A} R(p_A) |p_A,i_A\rangle_A\,
\mbox{}_A\langle p_A,i_A |.
\end{eqnarray}
%
%

We introduced the quantity 
%
%
\begin{equation}
R(p_A)=\frac{1}{Z}\sum_{p_B=0}^{N_B} 
\left(
\begin{array}{c}
N_B
\\
p_B
\end{array}
\right)
\mathrm{e}^{N\beta\left(\frac{p_A}{N}+\frac{p_B}{N}-\frac{1}{2}\right)^2},
\end{equation}
%
%
The partial entropy then simply reads
%
%
\begin{equation}
\mathcal{E}_{A}(\beta)=- \sum_{p_A=0}^{N_A}
\left(
\begin{array}{c}
N_A
\\
p_A
\end{array}
\right)
R(p_A) \log_2 R(p_A).
\end{equation}
%
%
Following the same line as for the partition function calculation and replacing
the binomials by the same form as (\ref{eq:bin_approx}), one can still use
Laplace's method to obtain the large-$N$ behavior of the partial entropy.
Nevertheless, things are a bit more involved since one now has to deal with a
double sum and thus with a two-variable integral. After some algebra, one gets
%
%
\begin{eqnarray}
\mathcal{E}_{A}(\beta<2) &=& \tau N-\frac{1}{2\log2}\bigg\{ \frac{\beta\: \tau}{2-\beta}  \\
&&+\log\bigg[\frac{2-\beta}{2-\beta(1-\tau)}\bigg]\bigg\}+O(1/N),\nonumber 
\end{eqnarray}
%
%
and
%
%
\begin{equation}
\label{eq:entrop_ana}
\mathcal{E}_{A}(\beta=2) = \tau N-\tau \sqrt{N}\frac{\sqrt{3}}{\log 2}\frac{\Gamma(3/4)}{\Gamma(1/4)}+\frac{1}{4}\log_2 N+O(N^0),
\end{equation}
%
%
where we have set $N_A/N=\tau$, and $N_B/N=1-\tau$. 

Finally, noting that $\mathcal{E}_{B}$ is obtained from $\mathcal{E}_{A}$ by
exchanging $\tau \leftrightarrow (1-\tau)$ and using (\ref{eq:entrop}), one obtains the following expressions of the mutual information
%
%
\begin{equation}
\label{eq:mi_ana}
\mathcal{I} (\beta<2)=  \frac{1}{2}\log_2 \bigg\{\frac{[2-\beta \: \tau][2-\beta
(1-\tau)]}{2(2-\beta)}\bigg\}+O(1/N),
\end{equation}
%
%
and
%
%
\begin{equation}
\label{eq:Icrit}
\mathcal{I} (\beta=2) = \frac{1}{4}\log_2 N+O(N^0).
\end{equation}
%
%

These expressions together with the exact numerical results for $\beta>2$ (in
the thermodynamical limit) are compared to finite-size numerical results in
Fig.~\ref{fig:mi_results} (left column).
In Fig.~\ref{fig:mi_results}(a), we display the mutual entropy as a function of
$T$ for several values of $N$ and $\tau=1/2$. The excellent agreement with
Eq.~(\ref{eq:mi_ana}) can be observed at large $N$.
In Fig.~\ref{fig:mi_results}(d) we provide a numerical check of
Eq.~(\ref{eq:Icrit}) that predicts a logarithmic divergence of the mutual
information at criticality, with a nontrivial prefactor  $1/4$. Note that in the present case ($h=0$), it is  possible to perform a numerical study for large system sizes (up to $N=2^{16}$ here) that allows one to observe how this divergence arises.
%

%
%
\subsection{Numerical results for $h>0$}
\label{subsec:results}
%
%

Let us now turn to the case $h\neq 0$ for which the spectrum, and hence $\mathcal{I}$ can only be computed numerically. Details of the algorithm we developed to compute $\mathcal{I}$ can be found in 
Appendix~\ref{sec:sub:algo}. 
In Figs.~\ref{fig:mi_results}(b) and (c), we display the behavior of the mutual
information as a function of temperature for $h=1/2$ and $h\simeq 0.9999$ (such
that $T_\mathrm{c}=0.1$), and for $\tau=1/2$.
For $h=1/2$, and as already observed for $h=0$ in Fig.~\ref{fig:mi_results}(a),
$\mathcal{I}$ reaches a finite value away from criticality whereas it increases
with the system size at $T_{\rm c}$.
Although this is less obvious for $h\simeq 0.9999$, it is however still true.
Indeed, the apparent increase of $\mathcal{I}$ away from the critical temperature observed
in Fig.~\ref{fig:mi_results}(c) is a finite-size artifact as can be inferred from the zero-temperature case. 
For $T=0$  and $0\leqslant h<1$, one has $\mathcal{I}=2\mathcal{E}_A-1$, since 
$\mathcal{E}_B=\mathcal{E}_A$ (for all $\tau$) and  $\mathcal{E}_{AB}=1$ in the broken phase ($\mathcal{E}_{AB}=0$ in the symmetric phase $h>1$). Then, using the expression of $\mathcal{E}_A$ computed in Refs.~\onlinecite{Barthel06_2,Vidal07}, one gets, in thermodynamical limit,
$\mathcal{I}\simeq 6.1113$ for $h\simeq 0.9999$ whereas for $N=512$ we find $\mathcal{I}\simeq 1.9378$. 
It is thus clear that, in this case, the asymptotic value is still far from being reached for $N=512$. 
By contrast, for $h=1/2$, one gets $\mathcal{I}\simeq 1.0286$ in the thermodynamic limit and  $\mathcal{I}\simeq 1.0291$ for $N=512$.

To investigate the divergence of the mutual information at criticality in more
details, we computed $\mathcal{I}(T_{\rm c})$ for increasing system sizes.
Results shown in Figs.~\ref{fig:mi_results}(d,e,f) confirm that
$\mathcal{I}(T_{\rm c})$ diverges logarithmically with $N$. 
Furthermore, it seems that the behavior of $\mathcal{I}$ for $0<h<1$ is the
same as for $h=0$. This can be seen in Fig.~\ref{fig:mi_N_all} where we have
superimposed data shown in Figs.~\ref{fig:mi_results}(d,e,f). Thus, although
numerically reachable sizes for $0<h <1$ (up to $N=2^{10}$ here) are not as
large as for $h=0$, we are led to conclude that $\mathcal{I}(T_{\rm c}) \sim \tfrac{1}{4} \log_2 N$ for all $0<h<1$. 
Actually, if one reminds that a similar result was found for the order parameter $m$ (see Fig.~\ref{fig:op_N}),
it is tempting to conjecture that, for all physical quantities, the leading finite-size corrections at finite temperatures can be infered from the zero-field problem. In other words, the system behaves classically for all nonvanishing temperatures. 

However, as already noted for the order parameter (see Fig.~\ref{fig:op_N}), things are dramatically different at zero temperature.  Indeed, in the purely quantum case (no thermal fluctuations),  a phase
transition occurs at $h=1$ and the mutual information can be computed
straightforwardly, as discussed above.  \\

\begin{figure*}
\begin{center}
\includegraphics[width=0.66\columnwidth]{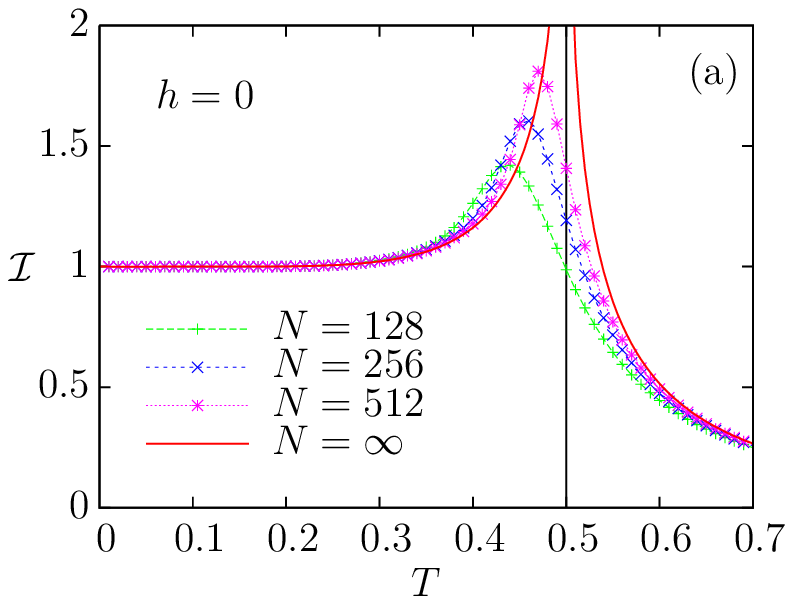}
\hfill
\includegraphics[width=0.66\columnwidth]{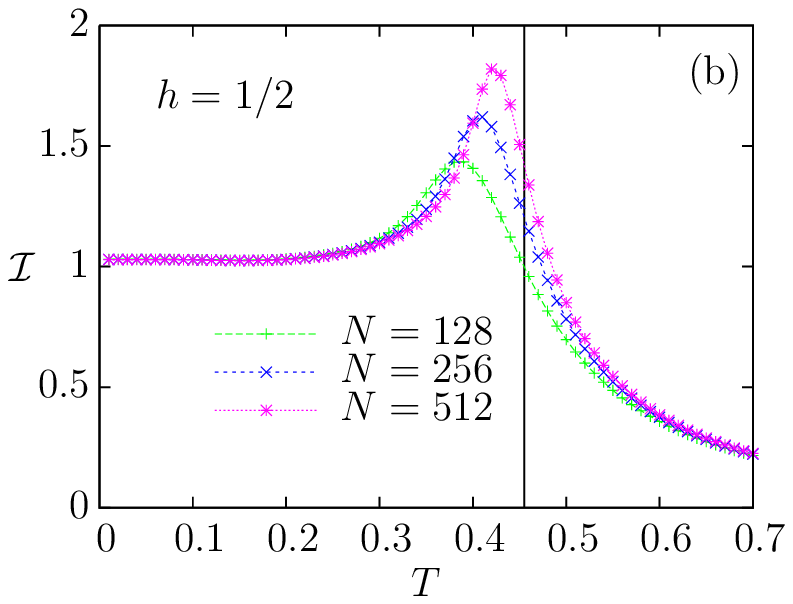}
\hfill
\includegraphics[width=0.66\columnwidth]{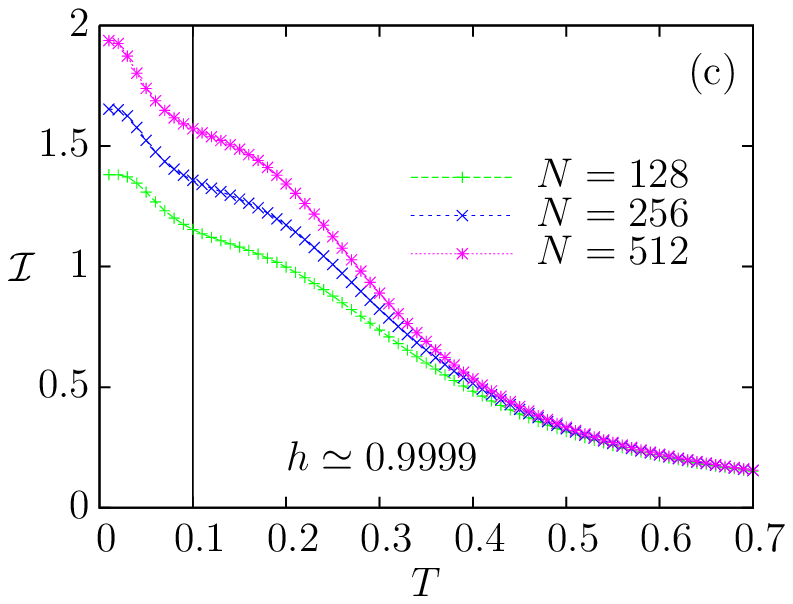}

\includegraphics[width=0.66\columnwidth]{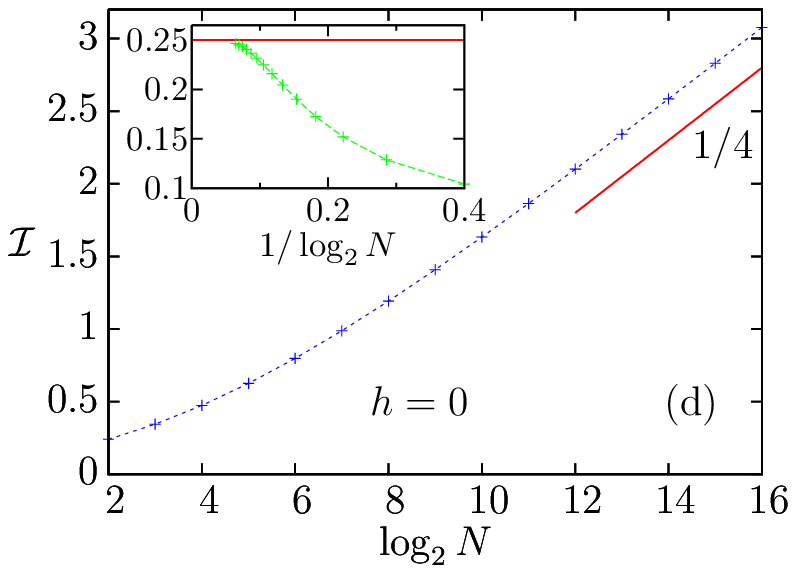}
\hfill
\includegraphics[width=0.66\columnwidth]{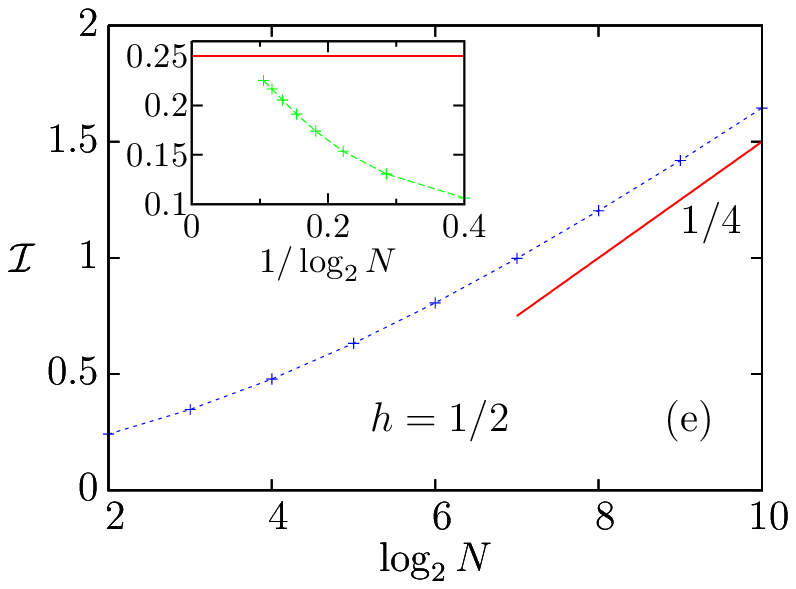}
\hfill
\includegraphics[width=0.66\columnwidth]{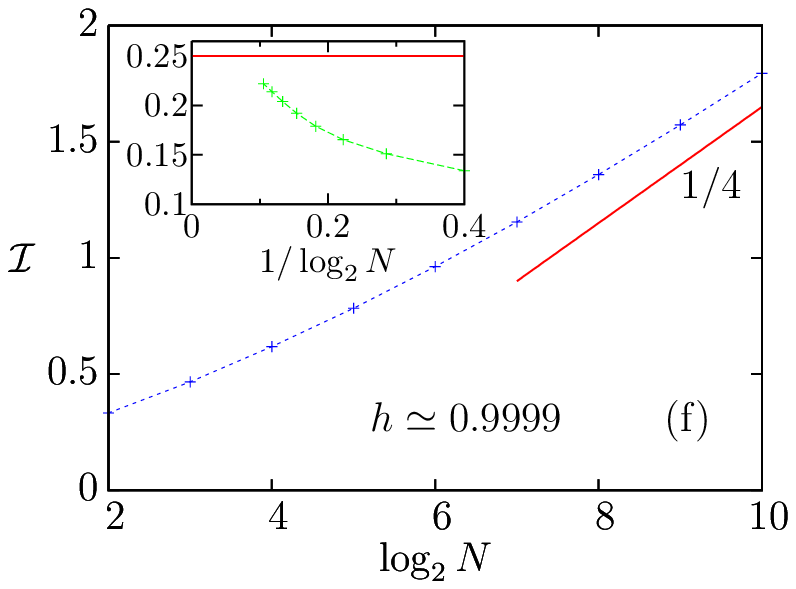}
\end{center}
\caption{Upper row : mutual information $\mathcal{I}$ as a function of $T$ for
$\tau=1/2$, various sizes, and for $h=0$ (a), $h=1/2$ (b) and $h\simeq 0.9999$
(c). The black vertical lines indicate the critical temperatures.
Lower row : scaling of the critical mutual information (evaluated at
$T=T_\mathrm{c}(h)$) as a function of $N$, for $\tau=1/2$, and for $h=0$ (d),
$h=1/2$ (e) and $h\simeq 0.9999$ (f).
The insets show the behaviour of the slope between two consecutive data points
as a function of $1/\log_2 N$. Red lines indicate the large-$N$ behavior
$\mathcal{I} \sim \tfrac{1}{4} \log_2 N$.}
\label{fig:mi_results}
\end{figure*}

Using the fact that  $\mathcal{E}_A\sim \tfrac{1}{6} \log_2 N$ for $h=1$ \cite{Barthel06_2,Vidal07}, and that  ${E}_{AB}=0$ at the critical point (the ground state is unique), one gets
%
%
\begin{equation}
\mathcal{I}(h=1,T=0) \sim \tfrac{1}{3}  \log_2 N,
\end{equation}
%
%
in contrast with the $\tfrac{1}{4} \log_2 N$ behavior found for \mbox{$0<h <1$}.

%
%
\begin{figure}[htp]
\includegraphics[width=0.8\columnwidth]{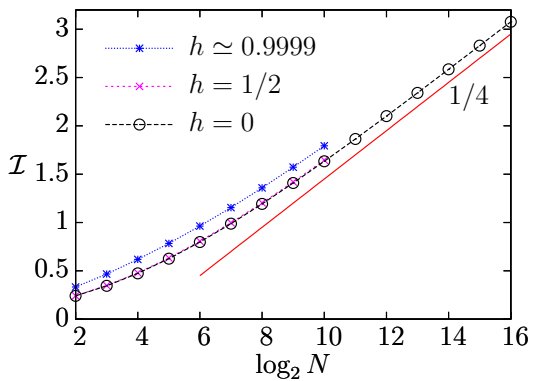}
\caption{Superposition of the curves of Figs.~\ref{fig:mi_results}(d,e,f).}
\label{fig:mi_N_all}
\end{figure}
%
%

%
%
\section{Conclusion}
%
%

We have studied the finite-temperature LMG model by focusing on
the mutual information $\mathcal{I}$ that measures both classical and quantum
many-body correlations in the system. 
As already pointed out in other recent studies \cite{Hastings10,Melko10,Singh11,Wilms11}, we
believe that mutual information is a good detector of finite-temperature phase
transitions. This quantity could turn out to be especially useful in systems
where no obvious or simple order parameter can be found (for instance in
topological phase transitions).
However, this quantity may reveal difficult to compute even in a
simple fully-connected model such as the one considered here. 

In the present work, we managed to get analytical results in the
simplest (purely classical) situation only but we obtained numerical
results in the whole parameter range for relatively large system sizes (up to $N=2^{10}$ spins).
As already observed for the (zero-temperature) entanglement entropy \cite{Barthel06_2,Vidal07}, we found
that $\mathcal{I}$ is always finite except on the transition line $T_{\rm c}(h)$
where it diverges logarithmically with the system size $N$. A complete study of the purely classical model allowed us to conjecture that for $0\leqslant h<1$ it diverges as $\frac{1}{4} \log_2 N$ whereas  for $h=1$, $\mathcal{I}$ behaves as $\frac{1}{3} \log_2 N$ at criticality. 

We hope that our results will motivate further studies of the mutual information (and more generally on many-body correlations) in related models since many questions remain open. First of all, let us underline that the model  studied here is  only a special case of the LMG model where spin-spin interactions are strongly anisotropic (Ising-like). Thus, investigating  the influence of an anisotropy parameter would be valuable especially since for isotropic interactions ($XX$-like), the full spectrum is known exactly (see for instance \cite{Dusuel05_2}) and, as already seen at zero temperature \cite{Vidal07},  one might expect a different behavior of $\mathcal{I}$ at the critical point. In addition, when competing ($XY$-like) interactions are present, there exists a special point in the parameter space where the ground-state is a separable state, the so-called Kurmann point \cite{Kurmann82}. It would be definitely instructive to analyze many-body correlations near this point at finite temperatures.

Second, natural extensions of fully-connected models have been recently analyzed
at zero temperature, revealing a rich variety of quantum phase transitions
\cite{Filippone11} among which some are of first order. The framework detailed
in App.~\ref{sec:sub:algo} clearly allows for a direct computation of the mutual
information in these models \cite{Wilms_note}. From that respect, it would also be  interesting to
study the finite-temperature Dicke model \cite{Dicke54} that shares many
entanglement features with the LMG model, as discussed in
Ref.~\onlinecite{Vidal07}. However, the presence of a bosonic mode coupled to a
set of two-level systems may give rise to interesting challenges if one aims at
computing the mutual information between the field and the atoms
(in particular
the procedure described in App.~\ref{sec:sub:algo} will not be sufficient).

\appendix

\section{General algorithm to compute the mutual information in spin-conserving Hamiltonian}
\label{sec:sub:algo}

The Hilbert space of a system consisting of $N$ spins 1/2 has dimension $2^N$
which usually limits the study of such systems to a number of spins of a few
tens. This restriction can be overcome in the LMG model, because its Hamiltonian
commutes with the total spin operator: $[H,\boldsymbol{S}^2]=0$. This symmetry
ensures that the Hamiltonian is block diagonal and that one can study each
block of fixed $s$ separately. The dimension of the latter is $2s+1$  (hence at
most linear in $N$) which allows one to study larger systems ($N\sim 10^3$ was
reached in this paper).

As is known from the theory of addition of angular momenta, there are $d_s^N$
distinct ways of obtaining a total spin $s$ when combining $N$ spins 1/2, where
$d_s^N$ reads
%
%
\begin{eqnarray}
   d_s^N &=& \begin{pmatrix}N\\N/2-s\end{pmatrix}
       -\begin{pmatrix}N\\N/2-s-1\end{pmatrix}\nonumber\\
   &=& \frac{2s+1}{N/2+s+1}\begin{pmatrix}N\\N/2-s\end{pmatrix}.
\end{eqnarray}
%
%
As a consequence, each energy level in the spin $s$-sector is (at least) $d_s^N$
times degenerate. Of course, one can check that
%
%
\begin{equation}
   \sum_{s=[S]}^{S}(2s+1)d_s^N=2^N,
\end{equation}
%
%
where $S=N/2$ is the maximum spin, and $[S]$ is the minimum spin, namely
$[S]=0$ if $N$ is even and $[S]=1/2$ if $N$ is odd. The fact that the sum over
$s$ runs over integers (half-odds) if $N$ is even (odd) is implicit.

The goal of this appendix is to explain how to use the spin-conserving property
to compute the reduced density matrix from which the mutual information is
extracted. Additional symmetries, such as the spin-flip symmetry
$[H,\prod_j\sigma_j^z]=0$ present in the LMG model, will not be considered
thereafter in order to give general results for any Hamiltonian satisfying
$[H,\boldsymbol{S}^2]=0$. Of course, additional symmetries can be
implemented to fasten the algorithm.

From the discussion given above, such a spin-conserving Hamiltonian can be block-diagonalized
and written as
%
%
\begin{eqnarray}
   H & = & \sum_{s=[S]}^S \, \sum_{i=1}^{d_s^N} H^{(s)}_i ,\\
     & = & \sum_{s=[S]}^S \, \sum_{i=1}^{d_s^N} \,
   \sum_{m=-s}^{s} \, \sum_{m'=-s}^{s}\, h^{(s)}_{m', m}\,
   \ket{s, m'}_i \,\, {}_i\bra{s, m}, \qquad
   \label{eq:def1}
\end{eqnarray}
%
%
where $i$ labels the $d_s^N$ degenerate subspaces of spin $s$, and where the
notations of the sums over $s$ have already been introduced above. Furthermore,
we introduced eigenstates $\ket{s, m}_i$ of operators $\mathbf{S}^2$ and $S_z$
with eigenvalues $s(s+1)$ and $m$ respectively. It is worth noting that matrix
elements $h^{(s)}_{m', m}$ are independent of $i$ so that, for each $s$, one has
$d_s^N$ copies of the same matrix to diagonalize. Once the diagonalizations are
performed, one can write
%
%
\begin{equation}
   H  = \sum_{s=[S]}^S \, \sum_{i=1}^{d_s^N} \,
   \sum_{\alpha=1}^{2s+1} E^{(s)}_{\alpha}
   \ket{s ; \alpha}_i \,\, {}_i\bra{s ; \alpha},
   \label{eq:diag}
\end{equation}
%
%
where eigenvalues $E^{(s)}_{\alpha}$ of $H^{(s)}_i$ are independent of $i$ and where the corresponding eigenvector $\ket{s ; \alpha}_i$ is given by
%
%
\begin{equation}
   \ket{s ; \alpha}_i = \sum_{m=-s}^{s} a_{\alpha ; m}^{(s)} \ket{s, m}_i,
   \label{eq:evec}
\end{equation}
%
%
with coefficients $a_{\alpha ; m}^{(s)} \in \mathbb{C}$ being independent of
$i$.

The partition function is then evaluated as
%
%
\begin{eqnarray}
   Z & = & \Tr \: {\rm e}^{-\beta H}, \\
   &=& \sum_{s=[S]}^S \, \sum_{i=1}^{d_s^N} \, \sum_{\alpha=1}^{2s+1} \rme^{-\beta E^{(s)}_{\alpha}} ,\\
   & = & \sum_{s=[S]}^S \, d_s^N \left[\sum_{\alpha=1}^{2s+1} \,
   \rme^{-\beta E^{(s)}_{\alpha}}\right]
   = \sum_{s=[S]}^S \, d_s^N Z^{(s)},
\end{eqnarray}
%
%
where $Z^{(s)}=\Tr\,\rme^{-\beta H^{(s)}_\mathrm{ref}}$ is the partition function associated to any of the $H^{(s)}_i$
(here we choose a reference index $i=\mathrm{ref}$).

In the same vein, one can write the entropy as
%
%
\begin{equation}
\mathcal{E}_{AB}  =  -\Tr [\rho\log_2\rho] = \sum_{s=[S]}^S \, d_s^N
   \mathcal{E}_{AB}^{(s)},
\end{equation}
%
%
where $\rho=\tfrac{1}{Z}\mathrm{e}^{-\beta H}$ is the density matrix and where
we defined $\mathcal{E}_{AB}^{(s)} =
-\Tr \left[\rho^{(s)}_\mathrm{ref}\log_2\rho^{(s)}_\mathrm{ref}\right]$
and
$\rho^{(s)}_\mathrm{ref} = \frac{1}{Z}\rme^{-\beta H^{(s)}_\mathrm{ref}}$.

%
%
\subsection*{Computation of $\mathcal{E}_A$ and $\mathcal{E}_B$}

We split the system in two subsystems $A$ and $B$ containing $L=\tau N$ and
$N-L=(1-\tau)N$ spins respectively, with the aim to compute the reduced density matrices $\rho_A=\Tr_B\rho$ and $\rho_B=\Tr_A\rho$.
The partial traces can be performed most easily by applying the technique
introduced above, for each subsystem. Therefore, one decomposes each spin
sector $s$ into spin subsector $s_1$ and $s_2$ (indices 1 and 2 refer to
subsystem $A$ and $B$ respectively). The Hamiltonian then reads
%
%
\begin{eqnarray}
   H &=& \sum_{s=[S]}^S \,
   \sum_{s_1=[S_1]}^{S_1}{\phantom\sum\hspace{-7mm}}' \,
   \sum_{s_2=[S_2]}^{S_2}{\phantom\sum\hspace{-7mm}}' \,
   \sum_{i_1=1}^{d_{s_1}^L} \,
   \sum_{i_2=1}^{d_{s_2}^{N-L}} H^{(s_1, s_2; s)}_{i_1, i_2}, \\
   &=&  \sum_{s=[S]}^S \,
   \sum_{s_1=[S_1]}^{S_1}{\phantom\sum\hspace{-7mm}}' \,
   \sum_{s_2=[S_2]}^{S_2}{\phantom\sum\hspace{-7mm}}' \,
   \sum_{i_1=1}^{d_{s_1}^L} \,
   \sum_{i_2=1}^{d_{s_2}^{N-L}}
   \sum_{m=-s}^{s} \, \sum_{m'=-s}^{s}\, \nonumber \\
   & & h^{(s)}_{m', m}\, \ket{s_1, s_2 ; s, m'}_{i_1, i_2}\,\, {}_{i_1, i_2}\bra{s_1, s_2 ; s, m},
\end{eqnarray}
%
%
where $\ket{s_1, s_2 ; s, m}_{i_1,i_2}$ denotes an eigenstate of operators
$\mathbf{S}_1^2, \mathbf{S}_2^2, \mathbf{S}^2$ and $S_z$ with eigenvalues
$s_1(s_1+1)$, $s_2(s_2+1)$, $s(s+1)$ and $m$ respectively. Index $i_1$ ($i_2$)
labels the $d_{s_1}^L$ ($d_{s_2}^{N-L}$) degenerate subspaces of spin $s_1$
($s_2$) that can be built from $L$ ($N-L$) spins 1/2.
For a given $s$, primed sums are restricted to values of $s_1$ and $s_2$ that
can add to a total spin $s$. That is to say, one must fulfill the inequalities
$|s_1-s_2|\leqslant s\leqslant s_1+s_2$, so one can alternatively write~:
%
%
\begin{equation}
   \sum_{s=[S]}^S \,
   \sum_{s_1=[S_1]}^{S_1}{\phantom\sum\hspace{-7mm}}' \,
   \sum_{s_2=[S_2]}^{S_2}{\phantom\sum\hspace{-7mm}}' \,
   = \sum_{s_1=[S_1]}^{S_1} \, \sum_{s_2=[S_2]}^{S_2} \,
   \sum_{s=|s_1-s_2|}^{s_1+s_2}.
\end{equation}
%
%
We have denoted \mbox{$S=N/2$}, $S_1=L/2$ and $S_2=(N-L)/2$ the maximum spins of
the whole system and of each subsystems. As before, minimum spins are denoted
with square brackets.
Note that the degeneracy $d_s^N$ of the spin-$s$ sector can be recovered from
%
%
\begin{equation}
   \sum_{s_1=[S_1]}^{S_1}{\phantom\sum\hspace{-7mm}}' \,
   \sum_{s_2=[S_2]}^{S_2}{\phantom\sum\hspace{-7mm}}' \,
   d_{s_1}^L d_{s_2}^{N-L} = d_s^N.
\end{equation}
%
%

The matrix elements $h^{(s)}_{m', m}$ are the same as in Eq.~(\ref{eq:def1}),
and do not depend on $s_1$, $s_2$, $i_1$ or $i_2$. Thus, all Hamiltonians
$H^{(s_1, s_2; s)}_{i_1, i_2}$ have the same eigenvalues $E^{(s)}_{\alpha}$
[which are the same as in Eq.~(\ref{eq:diag})], with the corresponding
eigenvectors
%
%
\begin{equation}
   \ket{s_1, s_2 ; s ; \alpha}_{i_1, i_2}=
   \sum_{m=-s}^{s} a_{\alpha ; m}^{(s)} \ket{s_1, s_2 ; s, m}_{i_1, i_2}. \\
\end{equation}
%
%
Once again, coefficients $a_{\alpha ; m}^{(s)} \in \mathbb{C}$ are independent
of $s_1$, $s_2$, $i_1$ and $i_2$, and are the same as in Eq.~(\ref{eq:evec}).

Then, the density matrix $\rho=\frac{1}{Z}\rme^{-\beta H}$ reads
%
%
\begin{widetext}
\begin{eqnarray}
   \rho & = &\frac{1}{Z} \sum_{s_1=[S_1]}^{S_1} \, \sum_{s_2=[S_2]}^{S_2} \,
       \sum_{s=|s_1-s_2|}^{s_1+s_2} \,
       \sum_{i_1=1}^{d_{s_1}^L} \, \sum_{i_2=1}^{d_{s_2}^{N-L}} \,
       \sum_{\alpha=1}^{2s+1} \,
       \rme^{-\beta E_\alpha^{(s)}}
       \ket{s_1, s_2 ; s ; \alpha}_{i_1, i_2} \,\,
       {}_{i_1, i_2}\bra{s_1, s_2 ; s ; \alpha} ,\\
   & = & \frac{1}{Z} \sum_{s_1=[S_1]}^{S_1} \, \sum_{s_2=[S_2]}^{S_2} \,
       \sum_{s=|s_1-s_2|}^{s_1+s_2} \,
       \sum_{i_1=1}^{d_{s_1}^L} \, \sum_{i_2=1}^{d_{s_2}^{N-L}} \,
       \sum_{\alpha=1}^{2s+1} \,
       \sum_{m=-s}^{s} \, \sum_{m'=-s}^{s}\,
       \rme^{-\beta E_\alpha^{(s)}}
       {a_{\alpha ; m}^{(s)}}^{\!\!\!*} a_{\alpha ; m'}^{(s)}
       \ket{s_1, s_2 ; s, m'}_{i_1, i_2} \,\,
       {}_{i_1, i_2}\bra{s_1, s_2 ; s, m}. \qquad
\end{eqnarray}
\end{widetext}
%
%
where $a^*$ denotes the complex conjugate of $a$.

One of the final steps is to decompose each basis state
$\ket{s_1, s_2 ; s, m'}_{i_1, i_2}$ on the tensor-product state basis using
the  Clebsch-Gordan coefficients
%
%
\begin{eqnarray}
   & & \ket{s_1, s_2 ; s, m'}_{i_1, i_2} =  \\
   & & \sum_{m_1=-s_1}^{s_1} \, \sum_{m_2=-s_2}^{s_2}
   C_{s_1, s_2 ; s}^{m_1, m_2 ; m}
   \ket{s_1, m_1}_{i_1} \otimes \ket{s_2, m_2}_{i_2}, \nonumber
\end{eqnarray}
%
%
with obvious notations. The only couples $(m_1, m_2)$ that contribute to the
above sum are such that $m_1+m_2=m$, since the Clebsch-Gordan coefficients
vanish otherwise.

Introducing the shorthand notation
%
%
\begin{widetext}
\begin{equation}
   \sum_{\mathrm{all}} \,=\,
       \sum_{s_1=[S_1]}^{S_1} \, \sum_{s_2=[S_2]}^{S_2} \,
       \sum_{s=|s_1-s_2|}^{s_1+s_2} \,
       \sum_{i_1=1}^{d_{s_1}^L} \, \sum_{i_2=1}^{d_{s_2}^{N-L}} \,
       \sum_{\alpha=1}^{2s+1} \,
       \sum_{m=-s}^{s} \, \sum_{m'=-s}^{s} \,
       \sum_{m_1=-s_1}^{s_1} \, \sum_{m_2=-s_2}^{s_2} \,
       \sum_{m_1'=-s_1}^{s_1} \, \sum_{m_2'=-s_2}^{s_2},
\end{equation}
%
%
one then gets
%
%
\begin{equation}
   \rho =\frac{1}{Z} \sum_{\mathrm{all}}
       \rme^{-\beta E_\alpha^{(s)}}
       {a_{\alpha ; m}^{(s)}}^{\!\!\!*} a_{\alpha ; m'}^{(s)}
       C_{s_1, s_2 ; s}^{m_1, m_2 ; m} C_{s_1, s_2 ; s}^{m_1', m_2' ; m'}
       \ket{s_1, m_1'}_{i_1} \,\, {}_{i_1}\bra{s_1, m_1} \otimes
       \ket{s_2, m_2'}_{i_2} \,\, {}_{i_2}\bra{s_2, m_2}.
\end{equation}
%
%
It is now possible to perform a partial trace. Let us focus on computing
$\rho_A=\Tr_B \rho$. This partial trace will enforce $m_2'=m_2$ in
$\sum_{\mathrm{all}}$. Furthermore, the index $i_2$ disappears from the quantity
to be summed over, so that $\sum_{i_2=1}^{d_{s_2}^{N-L}}$ simply yields a factor
$d_{s_2}^{N-L}$. At the end of the day, one finds
%
%
\begin{eqnarray}
   \rho_A & = &\frac{1}{Z}
       \sum_{s_1=[S_1]}^{S_1} \, \sum_{i_1=1}^{d_{s_1}^L} \,
       \sum_{m_1=-s_1}^{s_1} \, \sum_{m_1'=-s_1}^{s_1}
       {r_A}^{(s_1)}_{m_1', m_1}|s_1, m_1'\rangle_{i_1} \,\,
       {}_{i_1}\langle s_1, m_1| \quad \mbox{with}\nonumber\\
   {r_A}^{(s_1)}_{m_1', m_1} & = &
       \sum_{s_2=[S_2]}^{S_2} \,d_{s_2}^{N-L} \, \sum_{s=|s_1-s_2|}^{s_1+s_2} \,
       \sum_{m=\min_m}^{\max_m} \,
       \left(
       \sum_{\alpha=1}^{2s+1} \,
       \rme^{-\beta E_\alpha^{(s)}}
       {a_{\alpha ; m}^{(s)}}^{\!\!\!*} a_{\alpha ; m+m_1'-m_1}^{(s)}
       \right)
       C_{s_1, s_2 ; s}^{m_1, m-m_1 ; m} \,
       C_{s_1, s_2 ; s}^{m_1', m-m_1 ; m+m_1'-m_1}, \qquad
       \quad\label{eq:reduceddensity}
       \end{eqnarray}
\end{widetext}
%
%
with
%
%
\begin{eqnarray}
{\min}_m &=& \max(-s, -s+m_1-m_1', -s_2+m_1), \nonumber\\
{\max}_m &=& \min(s, s+m_1-m_1', s_2+m_1).
\end{eqnarray}
%
%
These limits come from the fact that all values of $m$ are not allowed
in between $-s$ and $s$, but one must also ensure that
\mbox{$-s_2 \leqslant m-m_1 \leqslant s_2$} and
\mbox{$-s \leqslant m+m_1'-m_1 \leqslant s$}.
Let us  remark that, here, we have got rid of the sum over $m_2$ and kept the
sum over~$m$. One could have of course done the opposite, resulting in a somewhat different
implementation.
Expression (\ref{eq:reduceddensity}) of $\rho_A$ allows one to compute
$\mathcal{E}_A$. The partial entropy $\mathcal{E}_B$ can be computed similarly.
Once these are known, the mutual information $\mathcal{I}$ is simply obtained
from its definition (\ref{eq:mutinf_def}).

It should be noted that the above calculation requires the numerical computation
of many Clebsch-Gordan coefficients. As there are too many of them, these cannot
be stored and must be computed when they are needed. This was achieved by
implementing the algorithm developed by Schulten and Gordon in Ref.~\onlinecite{Schulten75}, which is fast and stable.

As a final comment, let us give an estimate of the computational complexity of our
algorithm. Looking at Eq.~(\ref{eq:reduceddensity}), one sees that eight sums
are involved. However, the sum over $i_1$ will only yield a multiplicity, which
leaves seven sums. Furthermore, the values of the inner sum over $\alpha$,
namely
%
%
\begin{equation}
\sum_{\alpha=1}^{2s+1} \,
\rme^{-\beta E_\alpha^{(s)}}
{a_{\alpha ; m}^{(s)}}^{\!\!\!*} a_{\alpha ; m'}^{(s)},
\end{equation}
%
%
can be computed once (for a given $s$), and stored in a matrix. This matrix is
nothing but the density matrix in the spin-$s$ sector (up to a factor of
$1/Z$). The computation cost of this sum is thus negligible, and one is left
with a sum over six variables, namely $s_1$, $m_1$, $m_1'$, $s_2$, $s$ and $m$,
which  all take a number of values scaling linearly with $N$ (at most). So
we expect that our algorithm roughly scales as $N^6$ (which we observed in
practice).\\\\

\acknowledgments

This work is supported by the EU STReP QUE-VADIS,
the ERC grant QUERG, the FWF SFB grants FoQuS and
ViCoM, and the FWF Doctoral Programme CoQuS (W 1210).


\begin{thebibliography}{48}%
\makeatletter
\providecommand \@ifxundefined [1]{%
 \@ifx{#1\undefined}
}%
\providecommand \@ifnum [1]{%
 \ifnum #1\expandafter \@firstoftwo
 \else \expandafter \@secondoftwo
 \fi
}%
\providecommand \@ifx [1]{%
 \ifx #1\expandafter \@firstoftwo
 \else \expandafter \@secondoftwo
 \fi
}%
\providecommand \natexlab [1]{#1}%
\providecommand \enquote  [1]{``#1''}%
\providecommand \bibnamefont  [1]{#1}%
\providecommand \bibfnamefont [1]{#1}%
\providecommand \citenamefont [1]{#1}%
\providecommand \href@noop [0]{\@secondoftwo}%
\providecommand \href [0]{\begingroup \@sanitize@url \@href}%
\providecommand \@href[1]{\@@startlink{#1}\@@href}%
\providecommand \@@href[1]{\endgroup#1\@@endlink}%
\providecommand \@sanitize@url [0]{\catcode `\\12\catcode `\$12\catcode
  `\&12\catcode `\#12\catcode `\^12\catcode `\_12\catcode `\%12\relax}%
\providecommand \@@startlink[1]{}%
\providecommand \@@endlink[0]{}%
\providecommand \url  [0]{\begingroup\@sanitize@url \@url }%
\providecommand \@url [1]{\endgroup\@href {#1}{\urlprefix }}%
\providecommand \urlprefix  [0]{URL }%
\providecommand \Eprint [0]{\href }%
\providecommand \doibase [0]{http://dx.doi.org/}%
\providecommand \selectlanguage [0]{\@gobble}%
\providecommand \bibinfo  [0]{\@secondoftwo}%
\providecommand \bibfield  [0]{\@secondoftwo}%
\providecommand \translation [1]{[#1]}%
\providecommand \BibitemOpen [0]{}%
\providecommand \bibitemStop [0]{}%
\providecommand \bibitemNoStop [0]{.\EOS\space}%
\providecommand \EOS [0]{\spacefactor3000\relax}%
\providecommand \BibitemShut  [1]{\csname bibitem#1\endcsname}%
\let\auto@bib@innerbib\@empty
\bibitem [{\citenamefont {Eisert}\ \emph {et~al.}(2010)\citenamefont {Eisert},
  \citenamefont {Cramer},\ and\ \citenamefont {Plenio}}]{Eisert10}%
  \BibitemOpen
  \bibfield  {author} {\bibinfo {author} {\bibfnamefont {J.}~\bibnamefont
  {Eisert}}, \bibinfo {author} {\bibfnamefont {M.}~\bibnamefont {Cramer}}, \
  and\ \bibinfo {author} {\bibfnamefont {M.}~\bibnamefont {Plenio}},\ }\href
  {\doibase 10.1103/RevModPhys.82.277} {\bibfield  {journal} {\bibinfo
  {journal} {Rev. Mod. Phys.}\ }\textbf {\bibinfo {volume} {82}},\ \bibinfo
  {pages} {277} (\bibinfo {year} {2010})}\BibitemShut {NoStop}%
\bibitem [{\citenamefont {Holzhey}\ \emph {et~al.}(1994)\citenamefont
  {Holzhey}, \citenamefont {Larsen},\ and\ \citenamefont
  {Wilczek}}]{Holzhey94}%
  \BibitemOpen
  \bibfield  {author} {\bibinfo {author} {\bibfnamefont {C.}~\bibnamefont
  {Holzhey}}, \bibinfo {author} {\bibfnamefont {F.}~\bibnamefont {Larsen}}, \
  and\ \bibinfo {author} {\bibfnamefont {F.}~\bibnamefont {Wilczek}},\ }\href
  {\doibase 10.1016/0550-3213(94)90402-2} {\bibfield  {journal} {\bibinfo
  {journal} {Nucl. Phys. B}\ }\textbf {\bibinfo {volume} {424}},\ \bibinfo
  {pages} {443} (\bibinfo {year} {1994})}\BibitemShut {NoStop}%
\bibitem [{\citenamefont {Vidal}\ \emph {et~al.}(2003)\citenamefont {Vidal},
  \citenamefont {Latorre}, \citenamefont {Rico},\ and\ \citenamefont
  {Kitaev}}]{GVidal03}%
  \BibitemOpen
  \bibfield  {author} {\bibinfo {author} {\bibfnamefont {G.}~\bibnamefont
  {Vidal}}, \bibinfo {author} {\bibfnamefont {J.}~\bibnamefont {Latorre}},
  \bibinfo {author} {\bibfnamefont {E.}~\bibnamefont {Rico}}, \ and\ \bibinfo
  {author} {\bibfnamefont {A.}~\bibnamefont {Kitaev}},\ }\href {\doibase
  10.1103/PhysRevLett.90.227902} {\bibfield  {journal} {\bibinfo  {journal}
  {Phys. Rev. Lett.}\ }\textbf {\bibinfo {volume} {90}},\ \bibinfo {pages}
  {227902} (\bibinfo {year} {2003})}\BibitemShut {NoStop}%
\bibitem [{\citenamefont {Fidkowski}(2010)}]{Fidkowski10}%
  \BibitemOpen
  \bibfield  {author} {\bibinfo {author} {\bibfnamefont {L.}~\bibnamefont
  {Fidkowski}},\ }\href {\doibase 10.1103/PhysRevLett.104.130502} {\bibfield
  {journal} {\bibinfo  {journal} {Phys. Rev. Lett.}\ }\textbf {\bibinfo
  {volume} {104}},\ \bibinfo {pages} {130502} (\bibinfo {year}
  {2010})}\BibitemShut {NoStop}%
\bibitem [{\citenamefont {Turner}\ \emph {et~al.}(2010)\citenamefont {Turner},
  \citenamefont {Zhang},\ and\ \citenamefont {Vishwanath}}]{Turner10}%
  \BibitemOpen
  \bibfield  {author} {\bibinfo {author} {\bibfnamefont {A.~M.}\ \bibnamefont
  {Turner}}, \bibinfo {author} {\bibfnamefont {Y.}~\bibnamefont {Zhang}}, \
  and\ \bibinfo {author} {\bibfnamefont {A.}~\bibnamefont {Vishwanath}},\
  }\href {\doibase 10.1103/PhysRevB.82.241102} {\bibfield  {journal} {\bibinfo
  {journal} {Phys. Rev. B}\ }\textbf {\bibinfo {volume} {82}},\ \bibinfo
  {pages} {241102} (\bibinfo {year} {2010})}\BibitemShut {NoStop}%
\bibitem [{\citenamefont {Arnold}()}]{Arnold96}%
  \BibitemOpen
  \bibfield  {author} {\bibinfo {author} {\bibfnamefont {D.~V.}\ \bibnamefont
  {Arnold}},\ }\href@noop {} {}\bibinfo {note}
  {\href{http://www.complex-systems.com/abstracts/v10_i02_a03.html}{Complex
  Systems {\bf 10}, 143, (1996)}}\BibitemShut {NoStop}%
\bibitem [{\citenamefont {Sol\'e}\ \emph {et~al.}()\citenamefont {Sol\'e},
  \citenamefont {Manrubia}, \citenamefont {Luque}, \citenamefont {Delgado},\
  and\ \citenamefont {Bascompte}}]{Sole96}%
  \BibitemOpen
  \bibfield  {author} {\bibinfo {author} {\bibfnamefont {R.~V.}\ \bibnamefont
  {Sol\'e}}, \bibinfo {author} {\bibfnamefont {S.~C.}\ \bibnamefont
  {Manrubia}}, \bibinfo {author} {\bibfnamefont {B.}~\bibnamefont {Luque}},
  \bibinfo {author} {\bibfnamefont {J.}~\bibnamefont {Delgado}}, \ and\
  \bibinfo {author} {\bibfnamefont {J.}~\bibnamefont {Bascompte}},\ }\href@noop
  {} {}\bibinfo {note}
  {\href{http://193.146.190.110/~ricard/COMPLEXITY-96.pdf}{Complexity 13,
  (1996)}}\BibitemShut {NoStop}%
\bibitem [{\citenamefont {Matsuda}\ \emph {et~al.}(1996)\citenamefont
  {Matsuda}, \citenamefont {Kudo}, \citenamefont {Nakamura}, \citenamefont
  {Yamakawa},\ and\ \citenamefont {Murata}}]{Matsuda96}%
  \BibitemOpen
  \bibfield  {author} {\bibinfo {author} {\bibfnamefont {H.}~\bibnamefont
  {Matsuda}}, \bibinfo {author} {\bibfnamefont {K.}~\bibnamefont {Kudo}},
  \bibinfo {author} {\bibfnamefont {R.}~\bibnamefont {Nakamura}}, \bibinfo
  {author} {\bibfnamefont {O.}~\bibnamefont {Yamakawa}}, \ and\ \bibinfo
  {author} {\bibfnamefont {T.}~\bibnamefont {Murata}},\ }\href {\doibase
  10.1007/BF02330576} {\bibfield  {journal} {\bibinfo  {journal} {Int. J.
  Theor. Phys.}\ }\textbf {\bibinfo {volume} {35}},\ \bibinfo {pages} {839}
  (\bibinfo {year} {1996})}\BibitemShut {NoStop}%
\bibitem [{\citenamefont {Wilms}\ \emph {et~al.}()\citenamefont {Wilms},
  \citenamefont {Troyer},\ and\ \citenamefont {Verstraete}}]{Wilms11}%
  \BibitemOpen
  \bibfield  {author} {\bibinfo {author} {\bibfnamefont {J.}~\bibnamefont
  {Wilms}}, \bibinfo {author} {\bibfnamefont {M.}~\bibnamefont {Troyer}}, \
  and\ \bibinfo {author} {\bibfnamefont {F.}~\bibnamefont {Verstraete}},\
  }\href@noop {} {}\bibinfo {note}
  {\href{http://iopscience.iop.org/1742-5468/2011/10/P10011/}{J. Stat. Mech.
  (2011) P10011}}\BibitemShut {NoStop}%
\bibitem [{\citenamefont {Hu}\ \emph {et~al.}()\citenamefont {Hu},
  \citenamefont {Ronhovde},\ and\ \citenamefont {Nussinov}}]{Nussinov12}%
  \BibitemOpen
  \bibfield  {author} {\bibinfo {author} {\bibfnamefont {D.}~\bibnamefont
  {Hu}}, \bibinfo {author} {\bibfnamefont {P.}~\bibnamefont {Ronhovde}}, \ and\
  \bibinfo {author} {\bibfnamefont {Z.}~\bibnamefont {Nussinov}},\ }\href@noop
  {} {}\bibinfo {note}
  {\href{http://arxiv.org/abs/1008.2699}{arXiv:1008.2699}}\BibitemShut
  {NoStop}%
\bibitem [{\citenamefont {Lipkin}\ \emph {et~al.}(1965)\citenamefont {Lipkin},
  \citenamefont {Meshkov},\ and\ \citenamefont {Glick}}]{Lipkin65}%
  \BibitemOpen
  \bibfield  {author} {\bibinfo {author} {\bibfnamefont {H.~J.}\ \bibnamefont
  {Lipkin}}, \bibinfo {author} {\bibfnamefont {N.}~\bibnamefont {Meshkov}}, \
  and\ \bibinfo {author} {\bibfnamefont {A.~J.}\ \bibnamefont {Glick}},\ }\href
  {\doibase 10.1016/0029-5582(65)90862-X} {\bibfield  {journal} {\bibinfo
  {journal} {Nucl. Phys.}\ }\textbf {\bibinfo {volume} {62}},\ \bibinfo {pages}
  {188} (\bibinfo {year} {1965})}\BibitemShut {NoStop}%
\bibitem [{\citenamefont {Meshkov}\ \emph {et~al.}(1965)\citenamefont
  {Meshkov}, \citenamefont {Glick},\ and\ \citenamefont {Lipkin}}]{Meshkov65}%
  \BibitemOpen
  \bibfield  {author} {\bibinfo {author} {\bibfnamefont {N.}~\bibnamefont
  {Meshkov}}, \bibinfo {author} {\bibfnamefont {A.~J.}\ \bibnamefont {Glick}},
  \ and\ \bibinfo {author} {\bibfnamefont {H.~J.}\ \bibnamefont {Lipkin}},\
  }\href {\doibase 10.1016/0029-5582(65)90863-1} {\bibfield  {journal}
  {\bibinfo  {journal} {Nucl. Phys.}\ }\textbf {\bibinfo {volume} {62}},\
  \bibinfo {pages} {199} (\bibinfo {year} {1965})}\BibitemShut {NoStop}%
\bibitem [{\citenamefont {Glick}\ \emph {et~al.}(1965)\citenamefont {Glick},
  \citenamefont {Lipkin},\ and\ \citenamefont {Meshkov}}]{Glick65}%
  \BibitemOpen
  \bibfield  {author} {\bibinfo {author} {\bibfnamefont {A.~J.}\ \bibnamefont
  {Glick}}, \bibinfo {author} {\bibfnamefont {H.~J.}\ \bibnamefont {Lipkin}}, \
  and\ \bibinfo {author} {\bibfnamefont {N.}~\bibnamefont {Meshkov}},\ }\href
  {\doibase 10.1016/0029-5582(65)90864-3} {\bibfield  {journal} {\bibinfo
  {journal} {Nucl. Phys.}\ }\textbf {\bibinfo {volume} {62}},\ \bibinfo {pages}
  {211} (\bibinfo {year} {1965})}\BibitemShut {NoStop}%
\bibitem [{\citenamefont {Hastings}\ \emph {et~al.}(2010)\citenamefont
  {Hastings}, \citenamefont {Gonzalez}, \citenamefont {Kallin},\ and\
  \citenamefont {Melko}}]{Hastings10}%
  \BibitemOpen
  \bibfield  {author} {\bibinfo {author} {\bibfnamefont {M.~B.}\ \bibnamefont
  {Hastings}}, \bibinfo {author} {\bibfnamefont {I.}~\bibnamefont {Gonzalez}},
  \bibinfo {author} {\bibfnamefont {A.~B.}\ \bibnamefont {Kallin}}, \ and\
  \bibinfo {author} {\bibfnamefont {R.~G.}\ \bibnamefont {Melko}},\ }\href
  {\doibase 10.1103/PhysRevLett.104.157201} {\bibfield  {journal} {\bibinfo
  {journal} {Phys. Rev. Lett.}\ }\textbf {\bibinfo {volume} {104}},\ \bibinfo
  {pages} {157201} (\bibinfo {year} {2010})}\BibitemShut {NoStop}%
\bibitem [{\citenamefont {Melko}\ \emph {et~al.}(2010)\citenamefont {Melko},
  \citenamefont {Kallin},\ and\ \citenamefont {Hastings}}]{Melko10}%
  \BibitemOpen
  \bibfield  {author} {\bibinfo {author} {\bibfnamefont {R.~G.}\ \bibnamefont
  {Melko}}, \bibinfo {author} {\bibfnamefont {A.~B.}\ \bibnamefont {Kallin}}, \
  and\ \bibinfo {author} {\bibfnamefont {M.~B.}\ \bibnamefont {Hastings}},\
  }\href {\doibase 10.1103/PhysRevB.82.100409} {\bibfield  {journal} {\bibinfo
  {journal} {Phys. Rev. B}\ }\textbf {\bibinfo {volume} {82}},\ \bibinfo
  {pages} {100409} (\bibinfo {year} {2010})}\BibitemShut {NoStop}%
\bibitem [{\citenamefont {Singh}\ \emph {et~al.}(2011)\citenamefont {Singh},
  \citenamefont {Hastings}, \citenamefont {Kallin},\ and\ \citenamefont
  {Melko}}]{Singh11}%
  \BibitemOpen
  \bibfield  {author} {\bibinfo {author} {\bibfnamefont {R.~R.~P.}\
  \bibnamefont {Singh}}, \bibinfo {author} {\bibfnamefont {M.~B.}\ \bibnamefont
  {Hastings}}, \bibinfo {author} {\bibfnamefont {A.~B.}\ \bibnamefont
  {Kallin}}, \ and\ \bibinfo {author} {\bibfnamefont {R.~G.}\ \bibnamefont
  {Melko}},\ }\href {\doibase 10.1103/PhysRevLett.106.135701} {\bibfield
  {journal} {\bibinfo  {journal} {Phys. Rev. Lett.}\ }\textbf {\bibinfo
  {volume} {106}},\ \bibinfo {pages} {135701} (\bibinfo {year}
  {2011})}\BibitemShut {NoStop}%
\bibitem [{\citenamefont {Kallin}\ \emph {et~al.}(2011)\citenamefont {Kallin},
  \citenamefont {Hastings}, \citenamefont {Melko},\ and\ \citenamefont
  {Singh}}]{Kallin11}%
  \BibitemOpen
  \bibfield  {author} {\bibinfo {author} {\bibfnamefont {A.~B.}\ \bibnamefont
  {Kallin}}, \bibinfo {author} {\bibfnamefont {M.~B.}\ \bibnamefont
  {Hastings}}, \bibinfo {author} {\bibfnamefont {R.~G.}\ \bibnamefont {Melko}},
  \ and\ \bibinfo {author} {\bibfnamefont {R.~R.~P.}\ \bibnamefont {Singh}},\
  }\href {\doibase 10.1103/PhysRevB.84.165134} {\bibfield  {journal} {\bibinfo
  {journal} {Phys. Rev. B}\ }\textbf {\bibinfo {volume} {84}},\ \bibinfo
  {pages} {165134} (\bibinfo {year} {2011})}\BibitemShut {NoStop}%
\bibitem [{\citenamefont {Dutta}\ \emph {et~al.}()\citenamefont {Dutta},
  \citenamefont {Divakaran}, \citenamefont {Sen}, \citenamefont {Chakrabarti},
  \citenamefont {Rosenbaum},\ and\ \citenamefont {Aeppli}}]{Dutta11}%
  \BibitemOpen
  \bibfield  {author} {\bibinfo {author} {\bibfnamefont {A.}~\bibnamefont
  {Dutta}}, \bibinfo {author} {\bibfnamefont {U.}~\bibnamefont {Divakaran}},
  \bibinfo {author} {\bibfnamefont {D.}~\bibnamefont {Sen}}, \bibinfo {author}
  {\bibfnamefont {B.~K.}\ \bibnamefont {Chakrabarti}}, \bibinfo {author}
  {\bibfnamefont {T.~F.}\ \bibnamefont {Rosenbaum}}, \ and\ \bibinfo {author}
  {\bibfnamefont {G.}~\bibnamefont {Aeppli}},\ }\href@noop {} {}\bibinfo {note}
  {\href{http://arxiv.org/abs/1012.0653}{arXiv:1012.0653}}\BibitemShut
  {NoStop}%
\bibitem [{\citenamefont {Vidal}\ \emph {et~al.}(2004)\citenamefont {Vidal},
  \citenamefont {Palacios},\ and\ \citenamefont {Mosseri}}]{Vidal04_1}%
  \BibitemOpen
  \bibfield  {author} {\bibinfo {author} {\bibfnamefont {J.}~\bibnamefont
  {Vidal}}, \bibinfo {author} {\bibfnamefont {G.}~\bibnamefont {Palacios}}, \
  and\ \bibinfo {author} {\bibfnamefont {R.}~\bibnamefont {Mosseri}},\ }\href
  {\doibase 10.1103/PhysRevA.69.022107} {\bibfield  {journal} {\bibinfo
  {journal} {Phys. Rev. A}\ }\textbf {\bibinfo {volume} {69}},\ \bibinfo
  {pages} {022107} (\bibinfo {year} {2004})}\BibitemShut {NoStop}%
\bibitem [{\citenamefont {Dusuel}\ and\ \citenamefont
  {Vidal}(2004)}]{Dusuel04_3}%
  \BibitemOpen
  \bibfield  {author} {\bibinfo {author} {\bibfnamefont {S.}~\bibnamefont
  {Dusuel}}\ and\ \bibinfo {author} {\bibfnamefont {J.}~\bibnamefont {Vidal}},\
  }\href {\doibase 10.1103/PhysRevLett.93.237204} {\bibfield  {journal}
  {\bibinfo  {journal} {Phys. Rev. Lett.}\ }\textbf {\bibinfo {volume} {93}},\
  \bibinfo {pages} {237204} (\bibinfo {year} {2004})}\BibitemShut {NoStop}%
\bibitem [{\citenamefont {Dusuel}\ and\ \citenamefont
  {Vidal}(2005)}]{Dusuel05_2}%
  \BibitemOpen
  \bibfield  {author} {\bibinfo {author} {\bibfnamefont {S.}~\bibnamefont
  {Dusuel}}\ and\ \bibinfo {author} {\bibfnamefont {J.}~\bibnamefont {Vidal}},\
  }\href {\doibase 10.1103/PhysRevB.71.224420} {\bibfield  {journal} {\bibinfo
  {journal} {Phys. Rev. B}\ }\textbf {\bibinfo {volume} {71}},\ \bibinfo
  {pages} {224420} (\bibinfo {year} {2005})}\BibitemShut {NoStop}%
\bibitem [{\citenamefont {Latorre}\ \emph {et~al.}(2005)\citenamefont
  {Latorre}, \citenamefont {Or\'us}, \citenamefont {Rico},\ and\ \citenamefont
  {Vidal}}]{Latorre05_2}%
  \BibitemOpen
  \bibfield  {author} {\bibinfo {author} {\bibfnamefont {J.~I.}\ \bibnamefont
  {Latorre}}, \bibinfo {author} {\bibfnamefont {R.}~\bibnamefont {Or\'us}},
  \bibinfo {author} {\bibfnamefont {E.}~\bibnamefont {Rico}}, \ and\ \bibinfo
  {author} {\bibfnamefont {J.}~\bibnamefont {Vidal}},\ }\href {\doibase
  10.1103/PhysRevA.71.064101} {\bibfield  {journal} {\bibinfo  {journal} {Phys.
  Rev. A}\ }\textbf {\bibinfo {volume} {71}},\ \bibinfo {pages} {064101}
  (\bibinfo {year} {2005})}\BibitemShut {NoStop}%
\bibitem [{\citenamefont {Barthel}\ \emph {et~al.}(2006)\citenamefont
  {Barthel}, \citenamefont {Dusuel},\ and\ \citenamefont
  {Vidal}}]{Barthel06_2}%
  \BibitemOpen
  \bibfield  {author} {\bibinfo {author} {\bibfnamefont {T.}~\bibnamefont
  {Barthel}}, \bibinfo {author} {\bibfnamefont {S.}~\bibnamefont {Dusuel}}, \
  and\ \bibinfo {author} {\bibfnamefont {J.}~\bibnamefont {Vidal}},\ }\href
  {\doibase 10.1103/PhysRevLett.97.220402} {\bibfield  {journal} {\bibinfo
  {journal} {Phys. Rev. Lett.}\ }\textbf {\bibinfo {volume} {97}},\ \bibinfo
  {pages} {220402} (\bibinfo {year} {2006})}\BibitemShut {NoStop}%
\bibitem [{\citenamefont {Vidal}\ \emph {et~al.}()\citenamefont {Vidal},
  \citenamefont {Dusuel},\ and\ \citenamefont {Barthel}}]{Vidal07}%
  \BibitemOpen
  \bibfield  {author} {\bibinfo {author} {\bibfnamefont {J.}~\bibnamefont
  {Vidal}}, \bibinfo {author} {\bibfnamefont {S.}~\bibnamefont {Dusuel}}, \
  and\ \bibinfo {author} {\bibfnamefont {T.}~\bibnamefont {Barthel}},\
  }\href@noop {} {}\bibinfo {note}
  {\href{http://iopscience.iop.org/1742-5468/2007/01/P01015/}{J. Stat. Mech.
  (2007) P01015}}\BibitemShut {NoStop}%
\bibitem [{\citenamefont {Or\'us}\ \emph {et~al.}(2008)\citenamefont {Or\'us},
  \citenamefont {Dusuel},\ and\ \citenamefont {Vidal}}]{Orus08_2}%
  \BibitemOpen
  \bibfield  {author} {\bibinfo {author} {\bibfnamefont {R.}~\bibnamefont
  {Or\'us}}, \bibinfo {author} {\bibfnamefont {S.}~\bibnamefont {Dusuel}}, \
  and\ \bibinfo {author} {\bibfnamefont {J.}~\bibnamefont {Vidal}},\ }\href
  {\doibase 10.1103/PhysRevLett.101.025701} {\bibfield  {journal} {\bibinfo
  {journal} {Phys. Rev. Lett.}\ }\textbf {\bibinfo {volume} {101}},\ \bibinfo
  {pages} {025701} (\bibinfo {year} {2008})}\BibitemShut {NoStop}%
\bibitem [{\citenamefont {Wichterich}\ \emph {et~al.}(2010)\citenamefont
  {Wichterich}, \citenamefont {Vidal},\ and\ \citenamefont
  {Bose}}]{Wichterich10}%
  \BibitemOpen
  \bibfield  {author} {\bibinfo {author} {\bibfnamefont {H.}~\bibnamefont
  {Wichterich}}, \bibinfo {author} {\bibfnamefont {J.}~\bibnamefont {Vidal}}, \
  and\ \bibinfo {author} {\bibfnamefont {S.}~\bibnamefont {Bose}},\ }\href
  {\doibase 10.1103/PhysRevA.81.032311} {\bibfield  {journal} {\bibinfo
  {journal} {Phys. Rev. A}\ }\textbf {\bibinfo {volume} {81}},\ \bibinfo
  {pages} {032311} (\bibinfo {year} {2010})}\BibitemShut {NoStop}%
\bibitem [{\citenamefont {Ribeiro}\ \emph {et~al.}(2007)\citenamefont
  {Ribeiro}, \citenamefont {Vidal},\ and\ \citenamefont {Mosseri}}]{Ribeiro07}%
  \BibitemOpen
  \bibfield  {author} {\bibinfo {author} {\bibfnamefont {P.}~\bibnamefont
  {Ribeiro}}, \bibinfo {author} {\bibfnamefont {J.}~\bibnamefont {Vidal}}, \
  and\ \bibinfo {author} {\bibfnamefont {R.}~\bibnamefont {Mosseri}},\ }\href
  {\doibase 10.1103/PhysRevLett.99.050402} {\bibfield  {journal} {\bibinfo
  {journal} {Phys. Rev. Lett.}\ }\textbf {\bibinfo {volume} {99}},\ \bibinfo
  {pages} {050402} (\bibinfo {year} {2007})}\BibitemShut {NoStop}%
\bibitem [{\citenamefont {Ribeiro}\ \emph {et~al.}(2008)\citenamefont
  {Ribeiro}, \citenamefont {Vidal},\ and\ \citenamefont {Mosseri}}]{Ribeiro08}%
  \BibitemOpen
  \bibfield  {author} {\bibinfo {author} {\bibfnamefont {P.}~\bibnamefont
  {Ribeiro}}, \bibinfo {author} {\bibfnamefont {J.}~\bibnamefont {Vidal}}, \
  and\ \bibinfo {author} {\bibfnamefont {R.}~\bibnamefont {Mosseri}},\ }\href
  {\doibase 10.1103/PhysRevE.78.021106} {\bibfield  {journal} {\bibinfo
  {journal} {Phys. Rev. E}\ }\textbf {\bibinfo {volume} {78}},\ \bibinfo
  {pages} {021106} (\bibinfo {year} {2008})}\BibitemShut {NoStop}%
\bibitem [{\citenamefont {Quan}\ and\ \citenamefont
  {Cucchietti}(2009)}]{Quan09}%
  \BibitemOpen
  \bibfield  {author} {\bibinfo {author} {\bibfnamefont {H.~T.}\ \bibnamefont
  {Quan}}\ and\ \bibinfo {author} {\bibfnamefont {F.~M.}\ \bibnamefont
  {Cucchietti}},\ }\href {\doibase 10.1103/PhysRevE.79.031101} {\bibfield
  {journal} {\bibinfo  {journal} {Phys. Rev. E}\ }\textbf {\bibinfo {volume}
  {79}},\ \bibinfo {pages} {031101} (\bibinfo {year} {2009})}\BibitemShut
  {NoStop}%
\bibitem [{\citenamefont {Botet}\ and\ \citenamefont
  {Jullien}(1983)}]{Botet83}%
  \BibitemOpen
  \bibfield  {author} {\bibinfo {author} {\bibfnamefont {R.}~\bibnamefont
  {Botet}}\ and\ \bibinfo {author} {\bibfnamefont {R.}~\bibnamefont
  {Jullien}},\ }\href {\doibase 10.1103/PhysRevB.28.3955} {\bibfield  {journal}
  {\bibinfo  {journal} {Phys. Rev. B}\ }\textbf {\bibinfo {volume} {28}},\
  \bibinfo {pages} {3955} (\bibinfo {year} {1983})}\BibitemShut {NoStop}%
\bibitem [{\citenamefont {Liberti}\ \emph {et~al.}(2010)\citenamefont
  {Liberti}, \citenamefont {Piperno},\ and\ \citenamefont
  {Plastina}}]{Liberti10}%
  \BibitemOpen
  \bibfield  {author} {\bibinfo {author} {\bibfnamefont {G.}~\bibnamefont
  {Liberti}}, \bibinfo {author} {\bibfnamefont {F.}~\bibnamefont {Piperno}}, \
  and\ \bibinfo {author} {\bibfnamefont {F.}~\bibnamefont {Plastina}},\ }\href
  {\doibase 10.1103/PhysRevA.81.013818} {\bibfield  {journal} {\bibinfo
  {journal} {Phys. Rev. A}\ }\textbf {\bibinfo {volume} {81}},\ \bibinfo
  {pages} {013818} (\bibinfo {year} {2010})}\BibitemShut {NoStop}%
\bibitem [{\citenamefont {Kwok}\ \emph {et~al.}(2008)\citenamefont {Kwok},
  \citenamefont {Ning}, \citenamefont {Gu},\ and\ \citenamefont
  {Lin}}]{Kwok08}%
  \BibitemOpen
  \bibfield  {author} {\bibinfo {author} {\bibfnamefont {H.-M.}\ \bibnamefont
  {Kwok}}, \bibinfo {author} {\bibfnamefont {W.-Q.}\ \bibnamefont {Ning}},
  \bibinfo {author} {\bibfnamefont {S.-J.}\ \bibnamefont {Gu}}, \ and\ \bibinfo
  {author} {\bibfnamefont {H.-Q.}\ \bibnamefont {Lin}},\ }\href {\doibase
  10.1103/PhysRevE.78.032103} {\bibfield  {journal} {\bibinfo  {journal} {Phys.
  Rev. E}\ }\textbf {\bibinfo {volume} {78}},\ \bibinfo {pages} {032103}
  (\bibinfo {year} {2008})}\BibitemShut {NoStop}%
\bibitem [{\citenamefont {Ma}\ \emph {et~al.}(2008)\citenamefont {Ma},
  \citenamefont {Xu}, \citenamefont {Xiong},\ and\ \citenamefont
  {Wang}}]{Ma08}%
  \BibitemOpen
  \bibfield  {author} {\bibinfo {author} {\bibfnamefont {J.}~\bibnamefont
  {Ma}}, \bibinfo {author} {\bibfnamefont {L.}~\bibnamefont {Xu}}, \bibinfo
  {author} {\bibfnamefont {H.-N.}\ \bibnamefont {Xiong}}, \ and\ \bibinfo
  {author} {\bibfnamefont {X.}~\bibnamefont {Wang}},\ }\href {\doibase
  10.1103/PhysRevE.78.051126} {\bibfield  {journal} {\bibinfo  {journal} {Phys.
  Rev. E}\ }\textbf {\bibinfo {volume} {78}},\ \bibinfo {pages} {051126}
  (\bibinfo {year} {2008})}\BibitemShut {NoStop}%
\bibitem [{\citenamefont {Canosa}\ \emph {et~al.}(2007)\citenamefont {Canosa},
  \citenamefont {Matera},\ and\ \citenamefont {Rossignoli}}]{Canosa07}%
  \BibitemOpen
  \bibfield  {author} {\bibinfo {author} {\bibfnamefont {N.}~\bibnamefont
  {Canosa}}, \bibinfo {author} {\bibfnamefont {J.~M.}\ \bibnamefont {Matera}},
  \ and\ \bibinfo {author} {\bibfnamefont {R.}~\bibnamefont {Rossignoli}},\
  }\href {\doibase 10.1103/PhysRevA.76.022310} {\bibfield  {journal} {\bibinfo
  {journal} {Phys. Rev. A}\ }\textbf {\bibinfo {volume} {76}},\ \bibinfo
  {pages} {022310} (\bibinfo {year} {2007})}\BibitemShut {NoStop}%
\bibitem [{\citenamefont {Matera}\ \emph {et~al.}(2008)\citenamefont {Matera},
  \citenamefont {Rossignoli},\ and\ \citenamefont {Canosa}}]{Matera08}%
  \BibitemOpen
  \bibfield  {author} {\bibinfo {author} {\bibfnamefont {J.~M.}\ \bibnamefont
  {Matera}}, \bibinfo {author} {\bibfnamefont {R.}~\bibnamefont {Rossignoli}},
  \ and\ \bibinfo {author} {\bibfnamefont {N.}~\bibnamefont {Canosa}},\ }\href
  {\doibase 10.1103/PhysRevA.78.012316} {\bibfield  {journal} {\bibinfo
  {journal} {Phys. Rev. A}\ }\textbf {\bibinfo {volume} {78}},\ \bibinfo
  {pages} {012316} (\bibinfo {year} {2008})}\BibitemShut {NoStop}%
\bibitem [{\citenamefont {Amico}\ \emph {et~al.}(2008)\citenamefont {Amico},
  \citenamefont {Fazio}, \citenamefont {Osterloh},\ and\ \citenamefont
  {Vedral}}]{Amico08}%
  \BibitemOpen
  \bibfield  {author} {\bibinfo {author} {\bibfnamefont {L.}~\bibnamefont
  {Amico}}, \bibinfo {author} {\bibfnamefont {R.}~\bibnamefont {Fazio}},
  \bibinfo {author} {\bibfnamefont {A.}~\bibnamefont {Osterloh}}, \ and\
  \bibinfo {author} {\bibfnamefont {V.}~\bibnamefont {Vedral}},\ }\href
  {\doibase 10.1103/RevModPhys.80.517} {\bibfield  {journal} {\bibinfo
  {journal} {Rev. Mod. Phys.}\ }\textbf {\bibinfo {volume} {80}},\ \bibinfo
  {pages} {517} (\bibinfo {year} {2008})}\BibitemShut {NoStop}%
\bibitem [{\citenamefont {Wootters}(1998)}]{Wootters98}%
  \BibitemOpen
  \bibfield  {author} {\bibinfo {author} {\bibfnamefont {W.~K.}\ \bibnamefont
  {Wootters}},\ }\href {\doibase 10.1103/PhysRevLett.80.2245} {\bibfield
  {journal} {\bibinfo  {journal} {Phys. Rev. Lett.}\ }\textbf {\bibinfo
  {volume} {80}},\ \bibinfo {pages} {2245} (\bibinfo {year}
  {1998})}\BibitemShut {NoStop}%
\bibitem [{\citenamefont {Vidal}\ and\ \citenamefont {Werner}(2002)}]{Vidal02}%
  \BibitemOpen
  \bibfield  {author} {\bibinfo {author} {\bibfnamefont {G.}~\bibnamefont
  {Vidal}}\ and\ \bibinfo {author} {\bibfnamefont {R.~F.}\ \bibnamefont
  {Werner}},\ }\href {\doibase 10.1103/PhysRevA.65.032314} {\bibfield
  {journal} {\bibinfo  {journal} {Phys. Rev. A}\ }\textbf {\bibinfo {volume}
  {65}},\ \bibinfo {pages} {032314} (\bibinfo {year} {2002})}\BibitemShut
  {NoStop}%
\bibitem [{\citenamefont {Filippone}\ \emph {et~al.}(2011)\citenamefont
  {Filippone}, \citenamefont {Dusuel},\ and\ \citenamefont
  {Vidal}}]{Filippone11}%
  \BibitemOpen
  \bibfield  {author} {\bibinfo {author} {\bibfnamefont {M.}~\bibnamefont
  {Filippone}}, \bibinfo {author} {\bibfnamefont {S.}~\bibnamefont {Dusuel}}, \
  and\ \bibinfo {author} {\bibfnamefont {J.}~\bibnamefont {Vidal}},\ }\href
  {\doibase 10.1103/PhysRevA.83.022327} {\bibfield  {journal} {\bibinfo
  {journal} {Phys. Rev. A}\ }\textbf {\bibinfo {volume} {83}},\ \bibinfo
  {pages} {022327} (\bibinfo {year} {2011})}\BibitemShut {NoStop}%
\bibitem [{\citenamefont {K\"onig}\ \emph {et~al.}(2009)\citenamefont
  {K\"onig}, \citenamefont {Renner},\ and\ \citenamefont
  {Schaffner}}]{Koenig09}%
  \BibitemOpen
  \bibfield  {author} {\bibinfo {author} {\bibfnamefont {R.}~\bibnamefont
  {K\"onig}}, \bibinfo {author} {\bibfnamefont {R.}~\bibnamefont {Renner}}, \
  and\ \bibinfo {author} {\bibfnamefont {C.}~\bibnamefont {Schaffner}},\ }\href
  {\doibase 10.1109/TIT.2009.2025545} {\bibfield  {journal} {\bibinfo
  {journal} {IEEE Trans. Inf. Theory}\ }\textbf {\bibinfo {volume} {55}},\
  \bibinfo {pages} {4337} (\bibinfo {year} {2009})}\BibitemShut {NoStop}%
\bibitem [{\citenamefont {Ollivier}\ and\ \citenamefont
  {Zurek}(2001)}]{Ollivier01}%
  \BibitemOpen
  \bibfield  {author} {\bibinfo {author} {\bibfnamefont {H.}~\bibnamefont
  {Ollivier}}\ and\ \bibinfo {author} {\bibfnamefont {W.~H.}\ \bibnamefont
  {Zurek}},\ }\href {\doibase 10.1103/PhysRevLett.88.017901} {\bibfield
  {journal} {\bibinfo  {journal} {Phys. Rev. Lett.}\ }\textbf {\bibinfo
  {volume} {88}},\ \bibinfo {pages} {017901} (\bibinfo {year}
  {2001})}\BibitemShut {NoStop}%
\bibitem [{\citenamefont {Sarandy}(2009)}]{Sarandy09}%
  \BibitemOpen
  \bibfield  {author} {\bibinfo {author} {\bibfnamefont {M.~S.}\ \bibnamefont
  {Sarandy}},\ }\href {\doibase 10.1103/PhysRevA.80.022108} {\bibfield
  {journal} {\bibinfo  {journal} {Phys. Rev. A}\ }\textbf {\bibinfo {volume}
  {80}},\ \bibinfo {pages} {022108} (\bibinfo {year} {2009})}\BibitemShut
  {NoStop}%
\bibitem [{\citenamefont {Ferraro}\ \emph {et~al.}(2010)\citenamefont
  {Ferraro}, \citenamefont {Aolita}, \citenamefont {Cavalcanti}, \citenamefont
  {Cucchietti},\ and\ \citenamefont {Acin}}]{Ferraro09}%
  \BibitemOpen
  \bibfield  {author} {\bibinfo {author} {\bibfnamefont {A.}~\bibnamefont
  {Ferraro}}, \bibinfo {author} {\bibfnamefont {L.}~\bibnamefont {Aolita}},
  \bibinfo {author} {\bibfnamefont {D.}~\bibnamefont {Cavalcanti}}, \bibinfo
  {author} {\bibfnamefont {F.~M.}\ \bibnamefont {Cucchietti}}, \ and\ \bibinfo
  {author} {\bibfnamefont {A.}~\bibnamefont {Acin}},\ }\href {\doibase
  10.1103/PhysRevA.81.052318} {\bibfield  {journal} {\bibinfo  {journal} {Phys.
  Rev. A}\ }\textbf {\bibinfo {volume} {81}},\ \bibinfo {pages} {052318}
  (\bibinfo {year} {2010})}\BibitemShut {NoStop}%
\bibitem [{\citenamefont {Bender}\ and\ \citenamefont
  {Orszag}(1999)}]{Bender99}%
  \BibitemOpen
  \bibfield  {author} {\bibinfo {author} {\bibfnamefont {C.~M.}\ \bibnamefont
  {Bender}}\ and\ \bibinfo {author} {\bibfnamefont {S.~A.}\ \bibnamefont
  {Orszag}},\ }\href@noop {} {\emph {\bibinfo {title} {Advanced Mathematical
  Methods for Scientists and Engineers}}}\ (\bibinfo  {publisher}
  {Springer-Verlag},\ \bibinfo {address} {New-York},\ \bibinfo {year}
  {1999})\BibitemShut {NoStop}%
\bibitem [{\citenamefont {Kurmann}\ \emph {et~al.}(1982)\citenamefont
  {Kurmann}, \citenamefont {Thomas},\ and\ \citenamefont
  {M\"{u}ller}}]{Kurmann82}%
  \BibitemOpen
  \bibfield  {author} {\bibinfo {author} {\bibfnamefont {J.}~\bibnamefont
  {Kurmann}}, \bibinfo {author} {\bibfnamefont {H.}~\bibnamefont {Thomas}}, \
  and\ \bibinfo {author} {\bibfnamefont {G.}~\bibnamefont {M\"{u}ller}},\
  }\href {\doibase 10.1016/0378-4371(82)90217-5} {\bibfield  {journal}
  {\bibinfo  {journal} {Physica A}\ }\textbf {\bibinfo {volume} {112}},\
  \bibinfo {pages} {235} (\bibinfo {year} {1982})}\BibitemShut {NoStop}%
\bibitem [{\citenamefont {{J. Wilms {\it et al.}}}()}]{Wilms_note}%
  \BibitemOpen
  \bibfield  {author} {\bibinfo {author} {\bibnamefont {{J. Wilms {\it et
  al.}}}},\ }\href@noop {} {}\bibinfo {note} {{work in progress}}\BibitemShut
  {NoStop}%
\bibitem [{\citenamefont {Dicke}(1954)}]{Dicke54}%
  \BibitemOpen
  \bibfield  {author} {\bibinfo {author} {\bibfnamefont {R.~H.}\ \bibnamefont
  {Dicke}},\ }\href {\doibase 10.1103/PhysRev.93.99} {\bibfield  {journal}
  {\bibinfo  {journal} {Phys. Rev.}\ }\textbf {\bibinfo {volume} {93}},\
  \bibinfo {pages} {99} (\bibinfo {year} {1954})}\BibitemShut {NoStop}%
\bibitem [{\citenamefont {Schulten}\ and\ \citenamefont
  {Gordon}(1975)}]{Schulten75}%
  \BibitemOpen
  \bibfield  {author} {\bibinfo {author} {\bibfnamefont {K.}~\bibnamefont
  {Schulten}}\ and\ \bibinfo {author} {\bibfnamefont {R.~G.}\ \bibnamefont
  {Gordon}},\ }\href {\doibase 10.1063/1.522426} {\bibfield  {journal}
  {\bibinfo  {journal} {J. Math. Phys.}\ }\textbf {\bibinfo {volume} {16}},\
  \bibinfo {pages} {1961} (\bibinfo {year} {1975})}\BibitemShut {NoStop}%
\end{thebibliography}

%

\end{document}